\documentclass[preprint,aps]{revtex4}

\usepackage{graphicx}
\usepackage{dcolumn}
\usepackage{bm}
\usepackage{latexsym}
\usepackage{amsfonts}
\usepackage{amssymb}
\usepackage{amsmath}
\usepackage[usenames]{color}

\begin{document}
\preprint{KUNS-2414}

\title{Instability of Charged Lovelock Black Holes:\\Vector Perturbations and Scalar Perturbations}
\author{Tomohiro Takahashi}
\affiliation{Department of Physics,  Kyoto University, Kyoto, 606-8501, Japan
}

\date{\today}

\begin{abstract}
We examine the stability of charged Lovelock black hole solutions under vector type and scalar type perturbations.  
We find the suitable master variables for the stability analysis;  
the equations for these variables are the Schr${\ddot {\rm o}}$dinger type equations with two components and 
these Schr${\ddot {\rm o}}$dinger operators are symmetric.   
By these master equations, we show that charged Lovelock Black holes are stable under vector type perturbations.  
For scalar type perturbations, we show the criteria for the instability and check these numerically. 
In our previous paper, we have shown that nearly extremal black holes have the instability under tensor type perturbations. 
In this paper,  we find that  
black holes with small charge have the instability under scalar type perturbations even if they have relatively large mass. 
\end{abstract}

\pacs{98.80.Cq, 98.80.Hw}
\maketitle

\section{Introduction}
\label{section1}

The braneworld scenario with large extra dimensions predicts that 
higher dimensional black holes might be produced at colliders ~\cite{Giddings:2001bu}. 
Therefore,  higher dimensional black holes become attracting subjects and 
some aspects of these have been inspected so far. 
For example, exact solutions are investigated in higher dimensions. 
In higher dimensions, besides Schwarzschild black hole, Reissner-Nordstr${\rm{\ddot o}}$m black holes ~\cite{Tangherlini:1963bw} 
and rotating black holes~\cite{Myers:1986un},  
various solutions are found: black ring solution~\cite{Emparan:2001wk}, black di-ring~\cite{Iguchi:2007is}, black saturn~\cite{Elvang:2007rd}  and so on. 
For these solutions, from the standpoint of black hole creations, it is important 
to examine the stability of such solutions   
because stationary solutions with the instability  are not attractors of time evolution.  
This suggests that  such black holes  should not be realized.   

So far, various stability analyses for black hole solutions have been performed. 
One of the most notable analysis is that  
of Tangherlini-Schwarzschild solutions by Kodama and Ishibashi~\cite{Kodama:2003jz}.
They have derived master equations for all type perturbations. 
These are Shr${\ddot {\rm o}}$dinger type equations and 
they have shown that these Shr${\ddot {\rm o}}$dinger operators are all positive definite 
using the S-deformation approach which they have developed by Friedrichs extension. 
These results show that  Schwarzschild black holes are also stable in higher dimensions. 
They have also examine the stability of higher dimensional  Reissner-Nordstr${\rm{\ddot o}}$m black holes ~\cite{Kodama}. 
By the S-deformation, they have also shown that this charged solution is stable under tensor and vector type perturbations. 
For scalar type perturbations, it has been shown that this black hole is stable in 4 and 5 dimensions. 
For this  solution, the stability has also studied numerically and 
it has been found that black holes with large negative cosmological constants and large charge are unstable in more than 7-dimensions~\cite{Konoplya:2007jv}.
On Myers-Perry black hole solutions, in $D\geq 6$, 
there found the instability for singly rotating solutions when the spin parameter is large enough~\cite{Dias:2009iu}.   
In even dimensions, stability of near horizon geometry of rotating black hole with equal angular momenta are investigated~\cite{Tanahashi:2012si}.  
It has been suggested that scalar mode for base space has the instability. 
Recently, the stability of black ring solution is examined by using local Penrose inequality and 
it is shown that the fat branch is unstable~\cite{Figueras:2011he}. 

The stability analyses we have introduced above are all premised on Einstein theory. 
In fact, the stability of black hole solutions have been  examined mainly in Einstein theory. 
It is as important  as such analyses to investigate  the stability  in more general theories. 
In 4-dimensions, Einstein theory is characterized by two properties; 
the action has the general coordinate covariance and equation of motion  consists of metric, 
the first derivative of metric and the second derivative of metric~\cite{Lovelock:1972vz}. 
Then it is natural to extend the four dimensional gravitational theory to higher dimensional one keeping these two properties. 
In higher dimensions, the most general theory which satisfies above two features  is not Einstein theory; it is  Lovelock theory \cite{Lovelock:1971yv}. 
Then it is important to generalize the stability analysis of black hole solutions in Einstein theory  to these in Lovelock theory.  

Same as Einstein theory, a spherical symmetric solution is known  in Lovelock theory ~\cite{Christos,Wheeler:1985nh}.  
This solution is called as Lovelock black hole solution. 
For this Lovelock black hole solution, the stability has been analyzed in ~\cite{Dotti:2004sh,Gleiser:2005ra,Takahashi:2009dz}. 
In these papers, it has been shown that black holes with sufficiently small mass are unstable under scalar type perturbations in odd dimensions and 
unstable under tensor type perturbations in even dimensions. 
This critical mass differs with dimensions and Lovelock couplings. 
These instabilities become stronger as  wavelength becomes smaller, and the time scale of the instability converge to $0$ in small scale limit.  
Under vector perturbations, this solution is stable in all dimensions, which is independent of mass. 

Since black hole creations originate from protons at colliders,  
it is also important to take account of Maxwell-charge. In Lovelock theory with $U(1)$ field, 
a charged black hole solution is known~\cite{Christos}; this has spherical symmetry and a time-like Killing vector, and 
this Reissner-Nordstr${\rm{\ddot o}}$m like solution is called as charged Lovelock black hole solution.    
Then, in this paper, we'd like to  extend the stability analysis for Lovelock black hole solutions to charged Lovelock black hole solutions.  
For this charged solution, stability analysis under tensor type perturbations has been examined by us 
and we have shown that  black holes are unstable  if they have nearly extremal mass~\cite{Takahashi:2011qda}.  
In this paper, we extend our previous discussion to vector type perturbations  and scalar type perturbations; 
that is, we derive master equations for these type perturbations and examine the stability using master equations. 

The organization of this paper is as follows.
 In section \ref{section2}, we review Lovelock theory, present the charged Lovelock black hole solutions and 
 check the behavior of these solutions. We mainly concentrate on asymptotic flat branch.  
In section \ref{section3},  we review the analysis for tensor type perturbations~\cite{Takahashi:2011qda}.  
We examine tensor perturbations and show the criteria for stability under this type perturbations. 
In section \ref{section4}, we derive master equation for vector type perturbations and show that 
there is no instability under this type perturbation. 
In section \ref{section5},  we concentrate on scalar type perturbations. 
We show that master equations can be summarized as a Schr$\rm {\ddot o}$dinger type equation with two components, 
and using this equation we present  criteria for stability. 
In section \ref{section6},  we numerically examine the conditions for the instability  presented in section \ref{section3} and \ref{section5}. 
In this paper, we only check in $5-8$ dimensions. 
In the final section \ref{section7}, we summarize this paper. 
 
\section{Charged Lovelock Black Holes}
\label{section2}

In this section, we  introduce Lovelock theory and present  
charged black hole solutions in Lovelock-Maxwell theory. 
These solutions are expressed as the roots of the polynomial equation and 
we confirm that one of the roots is asymptotic flat.   
For this asymptotically flat root, we briefly check the behavior, singularity and horizons. 

\subsection{Lovelock-Maxwell System}
In Ref. \cite{Lovelock:1971yv}, 
D.Lovelock have constructed the gravitational theory whose equation of motion consists of 
 the  metric, the first derivative of the metric  and the second derivative of the metric.
The Lagrangian for this theory is 
 \begin{eqnarray}
  {\mathcal L}_{Lovelock}&=&-2\Lambda+\beta_1R\nonumber\\ 
  &\ &\  +\sum_{m=2}^{k} \frac{\beta_m(2m)!}{2^mm\prod_{p=1}^{2m-2}(n-p)} 
  \delta_{\kappa_1}^{[\lambda_1}\delta_{\rho_1}^{\sigma_1} \delta_{\kappa_2}^{\lambda_2}\cdots\delta_{\rho_m}^{\sigma_m]}
   R_{\lambda_1 \sigma_1}{}^{ \kappa_1\rho_1} \cdots  R_{\lambda_m \sigma_m}{}^{\kappa_m \rho_m}  \ ,   \nonumber
\end{eqnarray}
where $\Lambda$ corresponds to a cosmological constant and $\beta_m$s are arbitrary constants which we call Lovelock couplings. 
We add the coefficients $(2m)!/2^mm\prod_{p=1}^{2m-2}(n-p)$ for convenience. 
In the above Lagrangian, $n$ is related to dimension $D$ as $n=D-2$ and $k$ corresponds to the maximum order defined as $k\equiv [(D-1)/2]$ 
 where $[x]$ is the Gauss symbol. 
There exists maximum order $k$ due to the antisymmetric property of 
$\delta_{\kappa_1}^{[\lambda_1}\delta_{\rho_1}^{\sigma_1} \delta_{\kappa_2}^{\lambda_2}\cdots\delta_{\rho_m}^{\sigma_m]}$. 
When we fix the maximum order  $k$, 
the dimension is restricted as $n=D-2=2k-1,\ 2k$; for example, 
the second order Lovelock theory is the most general in $n=3$ or $n=4$, and 
the third order one is in $n=5$ or $n=6$. 
By the ambiguity of the overall factor of the action, we take the unit $\beta_1=1$ in this paper. 

In this paper, we want to  concentrate on Lovelock-Maxwell system. 
This system is described by the action  
\begin{eqnarray}
S=\int d^Dx \sqrt{-g}{\mathcal L}_{Lovelock}-\int d^Dx \sqrt{-g}\frac{1}{4}F_{\mu\nu}F^{\mu\nu},
\label{Lovelock_Maxwell_action}
\end{eqnarray}
where $F_{\mu\nu}$ is the field strength of Maxwell field $A_{\mu}$.  
In the action (\ref{Lovelock_Maxwell_action}), the dynamical variables  are $g_{\mu\nu}$ and $A_{\mu}$.  
The variations by these variables lead
\begin{eqnarray}
{\mathcal G}_{\mu}{}^{\nu}&=&T_{\mu}{}^{\nu}\label{Lovelock equations}\ ,\\
F^{\mu\nu}{}_{;\nu}&=&0 \label{maxwell}\ ,
\end{eqnarray}
where ${\mathcal G}_{\mu}{}^{\nu}$, which we call Lovelock tensor,  and $T_{\mu}{}^{\nu}$, which means energy momentum tensor for $U(1)$ field, 
are defined as 
\begin{eqnarray}
{\mathcal G}_{\mu}{}^{\nu}&=&\Lambda \delta_{\mu}^{\nu}+R_{\mu}{}^{\nu}-\frac{1}{2}R\delta_{\mu}{}^{\nu}\nonumber\\ 
&\ &\hspace{0.1cm}-\sum_{m=2}^{k}\frac{(2m+1)!}{2^{m+1}}\frac{\beta_m}{m\prod_{p=1}^{2m-2}(n-p)} 
	 \delta_{\mu}^{[\nu}\delta_{\kappa_1}^{\lambda_1}\delta_{\rho_1}^{\sigma_1} \delta_{\kappa_2}^{\lambda_2}\cdots\delta_{\rho_m}^{\sigma_m]}
       R_{\lambda_1 \sigma_1}{}^{\kappa_1\rho_1} \cdots  R_{\lambda_m \sigma_m}{}^{\kappa_m\rho_m}, \\
T_{\mu}{}^{\nu}&=&F_{\mu\lambda}F^{\nu\lambda}-\frac{1}{4}F_{\lambda\rho}F^{\lambda\rho}\delta_{\mu}^{\nu}\ .
\label{Maxwell_EMtensor}
\end{eqnarray} 
The field strength is defined as $F=dA$, then $F_{\mu\nu}$ must satisfy the identity   
\begin{eqnarray}
dF=0\ \Rightarrow F_{[\mu\nu;\lambda]}=0.
\label{maxwell_identity}
\end{eqnarray}
The above equations (\ref{Lovelock equations}), (\ref{maxwell}) and (\ref{maxwell_identity}) are our basic equations. 

\subsection{Charged Lovelock Black Holes}
For the basic equations, black hole solutions with two parameters are known~\cite{Christos}.  
We assume the static spherical symmetric metric with spherical symmetric electric field   
\begin{eqnarray}
ds^2&=&-f(r)dt^2+1/f(r)dr^2+r^2\gamma_{ij}dx^idx^j,
\label{metric_ansatz}\\
F^{tr}&=&E(r),\quad {\rm other\ components}=0\ .
\label{maxewll_ansatz} 
\end{eqnarray}
In these,  $\gamma_{ij}$ corresponds to the metric for $S^n$. 

We can easily check that these ansatz satisfies (\ref{maxwell_identity}). 
Then we concentrate on the others and these lead following equations; 
\begin{eqnarray}
F^{t\mu}{}_{;\mu}=0&\Rightarrow&\partial_r(r^nE(r))=0\ , \nonumber\\
{\mathcal G}_{i}{}^j=T_i{}^j&\Rightarrow& -\frac{1}{2r^{n-1}}\left(r^{n+1}{\mathcal P}[\psi]\right)^{\prime\prime}\delta_i^j=\frac{E^2}{2}\delta_i^j\ ,\nonumber\\
{\mathcal G}_t{}^t=T_t{}^t,\ {\mathcal G}_r{}^r=T_r{}^r&\Rightarrow& -\frac{n}{2r^n}\left(r^{n+1}{\mathcal P}[\psi]\right)^{\prime}=-\frac{E^2}{2}\ ,
\label{background_equations}
\end{eqnarray}
and the other components are identical. 
In (\ref{background_equations}), $\psi$ is related to $f(r)$ as $f=1-r^2\psi$ and ${\mathcal P}[\psi]$ is  defined as  
\begin{eqnarray}
	{\mathcal P}[\psi]\equiv\sum_{m=2}^{k}\left[\frac{\beta_m}{m}\psi^m\right]+\psi-\frac{2\Lambda}{n(n+1)}\ .
\end{eqnarray}
The second equation of (\ref{background_equations}) is derived by a derivative of the third equation with the first equation, 
so we only consider the first and third equations. 
The first equation can be easily integrated and the result is 
\begin{eqnarray}
E(r)=\sqrt{n(n-1)}{\mathcal Q}/r^n\ .
\label{maxewll_solution}
\end{eqnarray}
In this equation,  $\sqrt{n(n-1)}{\mathcal Q}$ is an integral constant  and  
this constant  corresponds  to the charge, which can be seen from the behavior of $E(r)$. 
Substituting (\ref{maxewll_solution}) into the third equation of (\ref{background_equations}), we can gain  
$\left(r^{n+1}{\mathcal P}[\psi]\right)^{\prime}=(n-1){\mathcal Q}^2/r^n$, or integrating both sides reads 
\begin{eqnarray} 
	 {\mathcal P}[\psi]=\frac{{\mathcal M}}{r^{n+1}}-\frac{{\mathcal Q}^2}{r^{2n}}\equiv M(r)\ ,\label{bg_polynomical_equation}
\end{eqnarray}
where ${\mathcal M}$ is an integral constant.  
We will see that ${\mathcal M}$ corresponds to mass when checking the asymptotic behavior of the solution~\cite{Myers:1988ze,Kofinas:2007ns}.

We must solve the polynomial equation (\ref{bg_polynomical_equation}) for solution of Lovelock-Maxwell system. 
In order to solve the polynomial equation (\ref{bg_polynomical_equation}),  
in this paper, we assume some conditions for Lovelock couplings $\beta_m$ and ${\mathcal M}$ for simplicity. 
First we consider mass of black hole is positive, that is, ${\mathcal M}>0$. 
Second we set cosmological constant $\Lambda=0$. 
In $\Lambda=0$, as we will see later, there must exist an asymptotic flat branch. 
Third, for simplicity, we assume the positivity of  Lovelock couplings, that is,  
\begin{eqnarray}
	\beta_m> 0\ (m\geq 2) \ . \label{assume_Lovelock_coefficients}
\end{eqnarray}

\subsection{Asymptotic Flat Branch}

Because (\ref{bg_polynomical_equation}) is $k$-th order polynomial,  the polynomial equation (\ref{bg_polynomical_equation}) should have at most $k$ solutions. 
However,  assuming all Lovelock couplings are positive  and  $\Lambda=0$, one of the roots corresponds to an asymptotic flat solution. 

For instance, we see the above statement when $k=2$. In this case, 
(\ref{bg_polynomical_equation}) reduces into the 2nd order polynomial equation, so this can be easily solved as 
\begin{eqnarray}
\psi(r)=
\left\{
\begin{array}{l}
(-1+\sqrt{1+2\beta_2{\mathcal M}/r^{n+1}-2\beta_2{\mathcal Q}^2/r^{2n}})/\beta_2\\
(-1-\sqrt{1+2\beta_2{\mathcal M}/r^{n+1}-2\beta_2{\mathcal Q}^2/r^{2n}})/\beta_2
\end{array}
\right.
\ .
\label{k2_psi_sol}
\end{eqnarray}
Let's consider the limit $r\rightarrow \infty$. The first root behaves as $\psi\rightarrow {\mathcal M}/r^{n+1}$ and the second  converges 
as $\psi\rightarrow -2/\beta_2$.  
Therefore, the function $f(r)=1-r^2\psi(r)$ behaves as $f=1-{\mathcal M}/r^{n-1}$ for the first root and $f=1+2r^2/\beta_2$ for the second one. 
This shows that the first branch expresses an asymptotic flat solution and the other is an asymptotic AdS solution. 

\begin{figure}[t]
 \begin{center}
  \includegraphics[width=90mm]{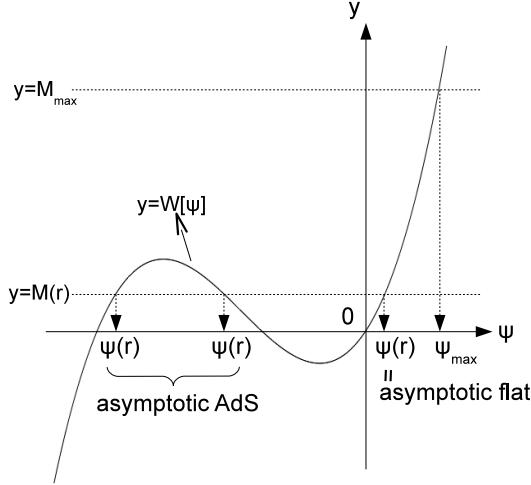}
 \end{center}
 \caption{We introduce the graphical method for finding the roots of  (\ref{bg_polynomical_equation}) in this figure.  
 The solid curve corresponds to $y={\mathcal P}[\psi]$ and the dotted horizontal line is $y=M(r)$ with fixed $r$. 
 The cross points in this figure are the solutions  of the  polynomial equation (\ref{bg_polynomical_equation}) for this $r$. 
 If we want to consider the roots for other radii, we draw the corresponding dotted line $y=M(r)$ 
 and see the cross points. }
 \label{fig_gr_method}
\end{figure}

Regrettably, we can not  write the roots of  (\ref{bg_polynomical_equation}) with general $k$ explicitly. 
Nevertheless, we can understand the existence of an asymptotic flat solution as the solution of (\ref{bg_polynomical_equation}).  
In order to check this statement, we want to introduce the graphical method. 
In Fig.\ref{fig_gr_method},  we  line $y=M(r)$ with fixed $r$ and $y={\mathcal P}[\psi]$ in $\psi-y$ diagram. 
The former is a horizontal line  because $M$ depends only on $r$ and we fix $r$. 
The cross points in Fig.\ref{fig_gr_method} are roots of the polynomial equation (\ref{bg_polynomical_equation}) for this fixed $r$, and   
if we want to find the roots for other radii, we move the horizontal line following the value of $M(r)$ and 
check the cross points. 

\begin{figure}[t]
 \begin{center}
  \includegraphics[width=100mm]{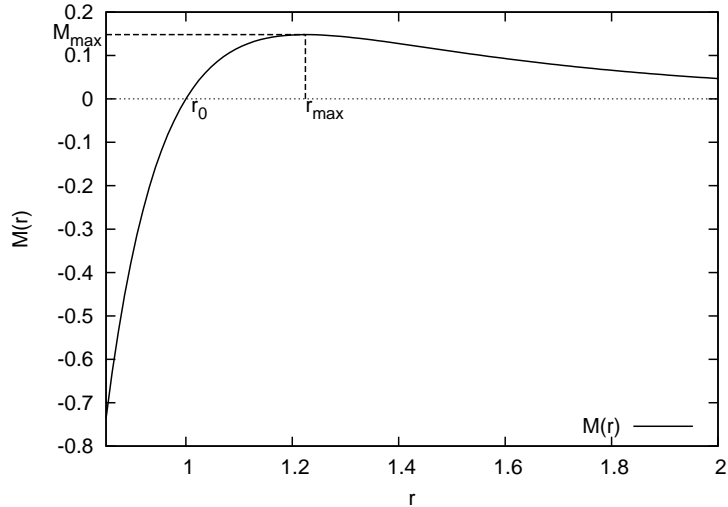}
 \end{center}
 \caption{We plot $M(r)$ in this figure with $n=3$, ${\mathcal M}=1$ and ${\mathcal Q}=1$. 
 $r_0$ and $r_{max}$ is defined as $M(r_0)=0$ and $M^{\prime}(r_{max})=0$ respectively. 
 $M_{max}$ corresponds  $M(r_{max})$. We can recognize the manners of roots of the polynomial equation (\ref{bg_polynomical_equation}) by this behavior and Fig.\ref{fig_gr_method}.}
 \label{fig_Mr}
\end{figure}

For this method, it is important to check the behavior of $M(r)={\mathcal M}/r^{n+1}-{\mathcal Q}^2/r^{2n}$.  
Its first derivative is $
M^{\prime}(r)=-\frac{(n+1){\mathcal M}}{r^{2n+1}}\left(r^{n-1}-\frac{2n{\mathcal Q}^2}{(n+1){\mathcal M}}\right)$. 
Then, like Fig.\ref{fig_Mr}, 
  $M(r)$ becomes $0$ at $r=r_0=\left({\mathcal Q}^2/{\mathcal M}\right)^{\frac{1}{n-1}}$, 
takes a maximum value $M_{max}=\frac{n-1}{2n}{\mathcal M}\left(\frac{(n+1){\mathcal M}}{2n{\mathcal Q}^2}\right)^{\frac{(n+1)}{(n-1)}}$at $r=r_{max=}\left(2n/(n+1)\right)^{\frac{1}{n-1}}r_0$ and behaves as  $M\sim {\mathcal M}/r^{n+1}$ in the asymptotic region. 

As mentioned above, $M(r)$ is positive in $r>r_0$. 
While $M(r)$ is positive, from Fig.\ref{fig_gr_method}, the polynomial equation (\ref{bg_polynomical_equation}) has only one positive root
 because ${\mathcal P}[\psi]$ satisfies ${\mathcal P}[0]=0$ and 
is a monotonically increasing function in $\psi>0$ under our assumptions (\ref{assume_Lovelock_coefficients}). 
This positive root expresses an asymptotic flat solution. 
To confirm this, let us consider the behavior of this $\psi$ when  $r\rightarrow \infty$. 
Because $M(r)$ converges to $0$ like ${\mathcal M}/r^{n+1}$ as $r\rightarrow\infty$, the positive cross point in Fig.\ref{fig_gr_method} also converges to $0$. 
In detail, eq.(\ref{bg_polynomical_equation}) with  $\psi\sim 0$ shows that  this root  converges like $\psi \sim {\mathcal M}/r^{n+1}$. 
Then, for this branch,  $f(r)$ behaves as $f(r)\sim 1-{\mathcal M}/r^{n-1}$ in the asymptotic region. 
The metric ansatz (\ref{metric_ansatz}) with this asymptotic behavior  shows  that this positive $\psi$ corresponds to an asymptotic flat solution. 
This asymptotic behavior of $f(r)$ also explains  that ${\mathcal M}$ corresponds to ADM mass.

In the last of this subsection, we  briefly check the behavior of our asymptotic flat root not in the asymptotic region. 
Let us consider by  the graphical method with Fig.\ref{fig_gr_method} again. 
Because $M(r)$ behaves as $0\rightarrow M_{max} \rightarrow 0$ when $r$ moves $\infty\rightarrow r_{max}\rightarrow r_0$,
our $\psi(r)$ varies as $0\rightarrow \psi_{max}\rightarrow 0$.  
When $r$ becomes smaller than $r_0$, $M(r)$ becomes negative and so 
$\psi(r)$ also takes negative values. 

\subsection{Singularities}

In $\psi<0$ or in $r<r_0$, our asymptotic flat solution has the curvature singularity.  
To confirm this,  we examine the  Kretschmann invariant  
\begin{eqnarray}
	R_{\mu \nu \lambda \rho}R^{\mu \nu \lambda \rho}=f^{\prime\prime}+2n\frac{f^{\prime2}}{r^2}+2n(n-1)\frac{(1-f)^2}{r^4}\ .\nonumber
\end{eqnarray}
This value diverges at $r=0$; and the singularity also exists  where $f^{\prime}$ or $f^{\prime\prime}$ diverge. 
For example,  $f^{\prime}$ has the term like  $r^2\psi^{\prime}$.  Form the the derivative of (\ref{bg_polynomical_equation}), this can be estimated as $r^2\psi^{\prime}=r^2M^{\prime}/\partial_{\psi} {\mathcal P}$.
Then, besides $r=0$, $R_{\mu\nu\lambda\rho}R^{\mu\nu\lambda\rho}$ also diverges where the derivative of ${\mathcal P}[\psi]$ with respect to $\psi$ becomes $0$. 
If  ${\mathcal P}[\psi]$ takes a extreme value at $\psi_0$, because ${\mathcal P}[\psi]$ is monotonically increasing in $\psi\geq 0$, 
such $\psi_0$ must be negative; that is, there is the singularity at $r_s(<r_0)$.  
If ${\mathcal P}[\psi]$ is monotonically incresing for all $\psi$, there is the curvature singularity at $r=0$. 
Therefore, whether ${\mathcal P}[\psi]$ has extreme values or not, our asymptotic flat branch has the curvature singularity somewhere in  $0\leq r<r_0$.

\subsection{Horizons}
Singularities must be wrapped by the event horizon from the standpoint of cosmic censorship. 
In this subsection, we consider  horizons and present the condition for existence of horizons. 

Our asymptotic flat solution has the event horizon at $f(r)=0$. 
This branch also satisfies ${\mathcal P}[\psi]=M(r)$, so  horizons can be determined from
\begin{eqnarray}
\left\{
\begin{array}{l}
0=1-r_H^2\psi_H\\
{\mathcal P}[\psi_H]=M(r_H)=\frac{{\mathcal M}}{r_H^{n+1}}-\frac{{\mathcal Q}^2}{r_H^{2n}}
\end{array}
\right.\ ,
\label{simultaneous}
\end{eqnarray}
where $r_H$  are horizon radii and  $\psi_H$ is defined as $\psi_H \equiv \psi(r_H) $. 
The first equation shows, if  horizons exist, the corresponding $\psi_H$ must be positive. 
Our $\psi(r)$ is positive in $r>r_0$, then  $r_H$, if exists, must satisfy $r_H>r_0$. 
As we have emphasized, the singularity exists somewhere in $r<r_0$. 
So we do not worry about the  naked singularity if (\ref{simultaneous}) has roots. 

\begin{figure}[t]
  \begin{center}
   \includegraphics[width=80mm]{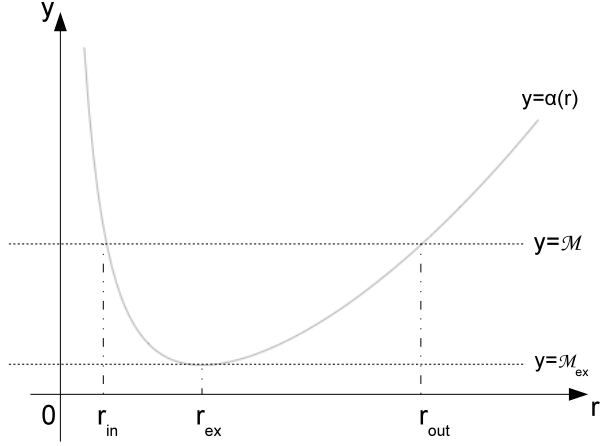}
  \end{center}
  \caption{The solid curve corresponds to $y=\alpha(r)$ and dotted line is $y={\mathcal M}$. In this figure, $y=\alpha(r)$ takes an extremal minimum at $r=r_{ex}$.  
  Because of (\ref{determinant_horizon}), the cross points mean horizon radii $r_H$.}
  \label{fig_alpha}
\end{figure}

Here, we consider the criteria for the existence of  roots of eq.(\ref{simultaneous}).
  Eliminating $\psi_H$ by  the first equation of (\ref{simultaneous}), the second equation becomes equation for $r _H$ as follows; 
 \begin{eqnarray}
&\ &{\mathcal P}\left[1/r_H^2\right]=\frac{{\mathcal M}}{r_H^{n+1}}-\frac{{\mathcal Q}^2}{r^{2n}_H}\nonumber\\ 
&\Leftrightarrow&{\mathcal M}=\frac{{\mathcal Q}^2}{r_H^{n-1}}+\left(r_H^{n-1}+\sum_{m=2}^k\frac{\beta_m}{m}r_H^{n+1-2m}\right)
\equiv \alpha(r_H)\ .
\label{determinant_horizon}
\end{eqnarray}  
The first term of $\alpha(r)$ is negative power of $r$ and its coefficient is positive. 
Under our assumption (\ref{assume_Lovelock_coefficients}), because $n=2k$ or $2k-1$, the other terms are positive power of $r$ and their coefficients are positive. 
Therefore, $y=\alpha(r)$ behaves as Fig.\ref{fig_alpha}; $\alpha(r)$ diverge near $r=0$, takes extremal minimum at $r=r_{ex}$ and monotonically 
increase in $r>r_{ex}$. 
Then, we can denote  that (\ref{determinant_horizon}) has two roots when ${\mathcal M}$ is larger than ${\mathcal M}_{ex}$ where  
\begin{eqnarray}
{\mathcal M}_{ex}\equiv\alpha(r_{ex})=\frac{{\mathcal Q}^2}{r_{ex}^{n-1}}+\left(r_{ex}^{n-1}+\sum_{m=2}^k\frac{\beta_m}{m}r_{ex}^{n+1-2m}\right)\ .
\label{determinant_extreme_mass}
\end{eqnarray}
In the two roots, the larger one corresponds to the outer horizon and we call this $r_{out}$ hereafter. 
Note that $\alpha(r)$ only depends on ${\mathcal Q}$ except for Lovelock couplings. 
Therefore, $r_{ex}$ is determined when we fix ${\mathcal Q}$; this shows that ${\mathcal M}_{ex}$ depends only on charge.  

Finally, we check the behaviors of some functions for the later discussions.  
First, we examine the behavior of $M(r)$ outside of $r_{out}$. 
The l.h.s of the first equation of (\ref{determinant_horizon}) is monotonically decreasing function and 
the r.h.s  is monotonically increasing in $r<r_{max}$ and monotonically decreasing in $r>r_{max}$. 
Then, when (\ref{determinant_horizon}) has two roots, it is forbidden that both of them are smaller than $r_{max}$; 
at least, the larger root $r_{out}$ must satisfy $r_{out}>r_{max}$. 
Hence, 
\begin{eqnarray}
M^{\prime}(r)<0\quad ({\rm r>r_{out}})\ .
\label{M_prime}
\end{eqnarray}
Then, in Fig.{\ref{fig_gr_method}}, while $r>r_{out}$ , the dotted line $y=M(r)$ falls monotonically as $r$ becomes larger 
and so $\psi(r)$ decreases monotonically in this region. 
Therefore, from the relation $\psi^{\prime}\partial_{\psi}{\mathcal P}=M^{\prime}$ reads 
\begin{eqnarray}
\partial_{\psi}{\mathcal P}[\psi]>0
\label{dW}
\end{eqnarray}
when we consider the outside of $r_{out}$.

 
\section{Tensor Type Perturbations}
\label{section3}

Thanks to the spherical symmetry of background (\ref{metric_ansatz}), 
tensor-type, vector-type and scalar-type perturbations are decomposed and 
we can examine them separately. 
We have already examined the tensor perturbations in \cite{Takahashi:2011qda}. 
In this section, we review our previous analysis briefly. 
Note that we only consider the case when there exist horizons; that is, ${\mathcal M}>{\mathcal M}_{ex}$ .

\subsection{Master Equation}
Under tensor type perturbation, there is no perturbation for Maxwell field. 
Then we only consider the gravitational perturbations 
\begin{eqnarray}
\delta g_{\mu\nu}=
\left(
\begin{array}{cc|c}
0&0&0\\
0&0&0\\
\hline
0&0&r^2\phi{\mathcal T}_{ij}
\end{array}
\right)
\ .
\end{eqnarray}
In this expressions, $\phi$ corresponds to the master variable. ${\mathcal T}_{ij}$ is the tensor harmonics which is characterized by  
the traceless condition  ${\mathcal T}^i{}_i=0$, 
the divergence-free condition ${\mathcal T}_{ij}{}^{|j}=0$ and the eigenequation ${\mathcal T}_{ij}{}^{|k}{}_{|k}=-(\ell(\ell+n-1)-2){\mathcal T}_{ij}$.  
Note that $|$ is the covariant derivative for $\gamma_{ij}$ and $\ell$ is the integer which satisfies $\ell\geq 2$ . 

Using above metric  perturbations, the first order equation $\delta {\mathcal G}_i{}^j=0$ leads~\cite{Takahashi:2009dz} 
\begin{eqnarray}
 T^{\prime}{\ddot \phi} - f^2T^{\prime}\phi^{\prime\prime}
      - f^2T^{\prime}\frac{(r^2fT^{\prime})^{\prime}}{r^2fT^{\prime}} \phi^{\prime}
      +  \frac{\ell(\ell+n-1)f}{(n-2)r}T^{\prime\prime} \phi 
      = 0\ ,
\label{tensor_perturbation_EOM}
\end{eqnarray}
where  $T(r)$ is  
\begin{eqnarray}
T(r)=r^{n-1}\partial_{\psi}{\mathcal P}[\psi], 
\end{eqnarray}
 which is  always positive in $r>r_{out}$ due to (\ref{dW}).

For this equation, as we have shown in \cite{Takahashi:2009dz,Takahashi:2011qda}, 
there exist ghost like instabilities if $T^{\prime}$ has negative regions. For example, 
the coefficient of  kinetic term in (\ref{tensor_perturbation_EOM}) is proportional to $T^{\prime}$, so 
this term has the wrong sing while $T^{\prime}$ is negative. 
Then, here we also assume $T^{\prime}(r)>0$ 
in $r>r_{out}$ for avoiding the ghost instability. 

Under  the ghost-free condition, we can change the normalization of $\phi$ as 
$
	\Psi(r)=\phi(r)r\sqrt{T^{\prime}(r)} 
$. 
Using this variable, converting  $r$ to $r^*$ which defined as  $dr^*/dr=1/f$ and Fourier transforming like $\Psi\rightarrow \Psi e^{i\omega t}$,
(\ref{tensor_perturbation_EOM}) is recast as 
\begin{eqnarray}
	{\mathcal H}\Psi=\omega^2\Psi \ , 
      \label{tensor_operator}
\end{eqnarray}
where
\begin{eqnarray}
	{\mathcal H}&=&-\partial_{r^*}^2+V_g(r)\nonumber\\
	V_g(r)&=&\frac{\ell(\ell+n-1)f}{(n-2)r}\frac{T^{\prime\prime}}{T^{\prime}}+\frac{1}{r\sqrt{T^{'}}}\partial_{r^*}^2(r\sqrt{T^{'}})\ . 
\end{eqnarray} 
Eq.(\ref{tensor_operator}) is a Schr${\rm {\ddot o}}$dinger type equation and its eigenvalue is $\omega^2$. 
Then, if this Schr${\rm {\ddot o}}$dinger operator ${\mathcal H}$ has negative spectra, 
we can say charged Lovelock black holes are unstable under tensor type perturbations. 

\subsection{Stability Analysis} 
In this subsection, we show ``there exist negative spectra if $T^{\prime\prime}$ has negative region in $r>r_{out}$" when $T^{\prime}$ is always positive.   
To show this, we define the inner product as 
\begin{eqnarray}
	(\chi_1,\chi_2)=\int_{-\infty}^{\infty} \chi_1^*\chi_2 dr^* \ ,\label{inner_product}
\end{eqnarray}
and use the inequality 
\begin{eqnarray}
	(\chi,{\mathcal H}\chi) \geq \omega_0^2\cdot(\chi,\chi)\ ,
      \label{spec_ineq}
\end{eqnarray}
where $\omega_0^2$ is the lower bound of spectra and $\chi$ is an arbitrary smooth function with  compact supports.  
From this inequality, we can show that there exist negative spectra  
if we find a function $\chi$ such that $(\chi,{\mathcal H}\chi)$ becomes negative under our  assumptions.

We assume $T^{\prime\prime}$ has negative regions and define $I$ 
as a closed set on such regions. 
Under these, we chose $\chi_0$ as a smooth function which has a compact support on $I$.  
For this $\chi_0$, $(\chi_0,{\mathcal H}\chi_0)$ is evaluated as 
\begin{eqnarray}
	(\chi_0,{\mathcal H}\chi_0)&=&\int_Idr^*\left[-\chi_0^*\partial_{r^*}^2\chi_0+V_g|\chi_0|^2\right]\nonumber\\
     &=&\int_{I} |\partial_{r^* }\chi_0-f\frac{d}{dr}\ln{(r\sqrt{T^{'}})} \chi_0|^2dr^{*}+\ell(\ell+n-1)
      \int_{I}\frac{f|\chi_0|^2}{(n-2)r}\frac{T^{\prime\prime}}{T^{\prime}}dr^* \ .
       \label{tensor_estimation}
\end{eqnarray}
In this calculation, we use Gauss divergence theorem and neglect boundary terms  
because $\chi_0$ is smoothly connecting to $0$ at $\partial I$. 
In (\ref{tensor_estimation}), the first term must be positive and the second integral is negative because 
we assume $T^{\prime}>0$ in $r>r_{out}$ and $T^{\prime\prime}<0$ on $I$.  
Therefore, taking $\ell\rightarrow \infty$, $(\chi_0,{\mathcal H}\chi_0)$ must become negative.   
Because of (\ref{spec_ineq}), this means negative spectra exist  in sufficiently large $\ell$ modes.  
Then we can declare that black holes are unstable if $T^{\prime\prime}$ takes negative values somewhere in $r>r_{out}$. 

Inversely, it can be also shown that charged Lovelock black holes are stable if $T^{\prime\prime}$ is always positive in $r\geq r_{out}$. 
As shown in Ref.~\cite{Kodama:2003jz}, it is sufficient for the stability to show that $(\Phi,{\mathcal H}\Phi)$ is positive for$\ ^{\forall}\Phi\in C_0^{\infty}(r^*)$.  
We can check this criterion by the same calculation of (\ref{tensor_estimation}) and the positivity of $T^{\prime\prime}$,  
so we can say black holes are stable if $T^{\prime\prime}$ is always positive. 
 
We want to summarize this section. 
For avoiding the ghost instability, we must assume $T^{\prime}$ is always positive. 
Under this assumption, charged Lovelock black holes are stable if and only if $T^{\prime\prime}$ always takes positive values in $r>r_{out}$. 
Hence, what we have to do is  a probe of  the behaviors of  $T^{\prime}$ and $T^{\prime\prime}$, and 
we will check these in the section \ref{section6}. 

In the end of this section, we want to comment about $T^{\prime}$ in the 2nd order Lovelock theory. 
We have already shown in our previous paper \cite{Takahashi:2011qda}, 
there is no ghost instability for this case. This can be checked by the direct calculations; 
eq.(\ref{k2_psi_sol}) leads $T(r)=r^{n-1}\partial_{\psi}{\mathcal P}[\psi]=r^{n-1}\sqrt{1+2\beta_2M(r)}$, then  $T^{\prime}$ is calculated as   
 \begin{eqnarray}
T^{\prime}=\frac{r^{n-2}}{\sqrt{1+2\beta_2M(r)}}\left[(n-1)+(n-3)\frac{\beta_2{\mathcal M}}{r^{n+1}}+2\frac{\beta_2{\mathcal Q}^2}{r^{2n}}\right]>0\ . 
\label{EGB_ghost}
\end{eqnarray}

\section{Vector Type Perturbations}
\label{section4}

In this section, we examine the vector type perturbations. 
We here also assume ${\mathcal M}>{\mathcal M}_{ex}$ for horizons and $T^{\prime}>0$ for no ghosts under tensor type perturbations. 
We only consider the perturbations in $r>r_{out}$. 
Under these assumptions, 
we derive master equations for vector type perturbations.   
Using this equations, we show that charged Lovelock black hole solutions are stable for vector type perturbations 
when $T^{\prime}$ is always positive. 

\subsection{Gravitational Perturbations}
Firstly, we'd like to consider the metric perturbations. 
In this paper, we use the Regge-Wheeler gauge in which metric perturbations are expressed as 
\begin{eqnarray}
\delta g_{\mu\nu}=
\left(
\begin{array}{cc|c}
0&0&h_1{\mathcal V}_i\\
0&0&h_2{\mathcal V}_i\\
\hline
 sym&sym&{\bf 0}
\end{array}
\right)\ .
\label{deltag_vector}
\end{eqnarray}
In these,  ${\mathcal V}_i$ is the vector harmonics which is characterized by the transverse condition ${\mathcal V}_i{}^{|i}=0$ and the 
eigenequation ${\mathcal V}_{i}{}^{|l}{}_{|l}=-\kappa_v {\mathcal V}_i$ with $\kappa_v=\ell(\ell+n-1)-1$ $(\ell\geq 1)$.

For vector type perturbations, other than $\delta{\mathcal G}_{t}{}^i$, $\delta{\mathcal G}_{r}{}^i$ and $\delta{\mathcal G}_{i}{}^j$ are trivial. 
From the above metric, we can calculate the non-trivial components as
\begin{eqnarray}
\delta {\mathcal G}_t{}^i&=&\left[\frac{(\kappa_v-(n-1))T^{\prime}}{2(n-1)r^{n+2}}h_1-\frac{f}{2r^{n+2}}\left\{r^3T\left(\left(\frac{h_1}{r^2}\right)^{\prime}-\frac{{\dot h_2}}{r^2}\right)\right\}^{\prime}+\frac{E^2}{r^2}h_1\right]{\mathcal V}^i\ ,\nonumber\\
\delta {\mathcal G}_r{}^i&=&\left[\frac{(\kappa_v-(n-1))T^{\prime}}{2(n-1)r^{n+2}}h_2-\frac{1}{2r^{n+2}}\frac{r^3T}{f}\left\{\left(\frac{h_1}{r^2}\right)^{\prime}-\frac{{\dot h_2}}{r^2}\right\}^{\cdot}+\frac{E^2}{r^2}h_2\right]{\mathcal V}^i\ ,\nonumber\\
\delta{\mathcal G}_i{}^j&=&\frac{1}{2(n-1)}\left[-\frac{T^{\prime}}{f}{\dot h_1}+\left(fT^{\prime}h_2\right)^{\prime}\right]\left({\mathcal V}_i{}^{|j}+{\mathcal V}_j{}^{|i}\right)\ . 
\label{vector_Lovelock}
\end{eqnarray}
This is almost same as the results of~\cite{Takahashi:2009dz} except for the background electric field. 
This gap mainly arises from the difference of the identity 
\begin{eqnarray}
\left\{(rf^{\prime}+2(1-f))T\right\}^{\prime}=2\frac{(r^nE)^2}{r^{n}}\left(\neq 0\right)\ .\nonumber
\label{}
\end{eqnarray}

\subsection{Perturbation of Maxwell field}
Next we examine the vector perturbations of Maxwell field. We start from the perturbation of vector potential
\begin{eqnarray}
\delta A_{\mu}=\left(0,\ 0,\ C{\mathcal V}_i\right)^T\ ,
\end{eqnarray}
where ${\mathcal V}_i$ is the vector harmonics. Note that $C$ is gauge invariant under $U(1)$ gauge because  there is no gauge freedom for vector perturbations. 
Using above $\delta A_{\mu}$, we can easily calculate the first order of the field strength as 
\begin{eqnarray}
\delta F_{ti}={\dot C}{\mathcal V}_{i}, \ \delta F_{ri}=C^{\prime}{\mathcal V}_i,\ \delta F_{ij}=C\left({\mathcal V}_{j|i}-{\mathcal V}_{i|j}\right),\ {\rm otherwise}=0\ .
\label{maxwell_per_vector}
\end{eqnarray}

Here, we derive the evolution equation for first order variable $C$ from Maxwell equations. 
It is easy to check the above field strength satisfies the identity $\delta F_{[\mu\nu;\lambda]}=0$. 
Therefore, $\delta(F_{\mu\nu}{}^{;\nu})=0$ is important for the evolution equation. 
In these equations, $\mu=t,r$ components are trivial and $\mu=i$ components read
\begin{eqnarray}
\frac{1}{f}{\ddot C}-\frac{1}{r^{n-2}}\partial_{r}\left(r^{n-2}f\partial_r C\right)-r^2E\left(\left(\frac{h_1}{r^2}\right)^{\prime}-\frac{{\dot h_2}}{r^2}\right)
+\frac{\kappa_v+(n-1)}{r^2}C=0\ .
\label{vector_maxwell_eq}
\end{eqnarray}

In order to gain evolution equations for gravitational field, we must calculate the first order of the energy momentum tensor (\ref{Maxwell_EMtensor}). 
This tensor consists of $\delta F_{\mu\nu}$, $\delta g_{\mu\nu}$ and background variables, 
so this tensor can be calculated from (\ref{maxwell_per_vector}), (\ref{deltag_vector}), (\ref{metric_ansatz}) and (\ref{maxewll_ansatz}), and the results are as follows;
\begin{eqnarray}
\delta T_t{}^i&=&\left[\frac{E^2}{r^2}h_1+\frac{Ef}{r^2}C^{\prime}\right]{\mathcal V}^i\ ,\nonumber\\
\delta T_r{}^i&=&\left[\frac{E^2}{r^2}h_2+\frac{E}{fr^2}{\dot C}\right]{\mathcal V}^i\ , \nonumber\\
{\rm other\ components}&=&0\ .
\label{vector_EM}
\end{eqnarray}
Then, from (\ref{vector_Lovelock}) and (\ref{vector_EM}), the first order Lovelock equation $\delta{\mathcal G}_{\mu}{}^{\nu}=\delta T_{\mu}{}^{\nu}$ reads 
\begin{eqnarray}
&\ &\frac{T^{\prime}}{2(n-1)r^{n+2}}\left(\kappa_v-(n-1)\right)h_1-\frac{f}{2r^{n+2}}\left\{r^3T\left(\left(\frac{h_1}{r^2}\right)^{\prime}-\frac{{\dot h_2}}{r^2}\right)\right\}^{\prime}=\frac{Ef}{r^2}C^{\prime}\ ,\nonumber\\
&\ &\frac{T^{\prime}}{2(n-1)r^{n+2}}\left(\kappa_v-(n-1)\right)h_2-\frac{1}{2r^{n+2}}\frac{1}{f}\left\{r^3T\left(\left(\frac{h_1}{r^2}\right)^{\prime}-\frac{{\dot h_2}}{r^2}\right)\right\}^{\cdot}=\frac{E}{fr^2}{\dot C}\ ,\nonumber\\
&\ &-\frac{T^{\prime}}{f}{\dot h_1}+\left(fT^{\prime}h_2\right)^{\prime}=0\ .
\label{vector_EOM}
\end{eqnarray}

\subsection{Master Equations}
Now we are position to derive master equations from (\ref{vector_EOM}) and (\ref{vector_maxwell_eq}). 
First of all, we treat the third equation of (\ref{vector_EOM}). From this equation we can define a new variable $\phi$ as 
\begin{eqnarray}
h_1=\frac{f}{T^{\prime}}\phi^{\prime}\ , \quad h_2=\frac{1}{fT^{\prime}}{\dot \phi}\ .
\label{vector_subst}
\end{eqnarray}
Substituting (\ref{vector_subst}) into the first equation of (\ref{vector_EOM}) and integrating this with respect to $r$ reads 
\begin{eqnarray}
2(Er^n)C+C_1(t)=\frac{\kappa_v-(n-1)}{n-1}\phi-r^3T\left(\left(\frac{h_1}{r^2}\right)^{\prime}-\frac{{\dot h_2}}{r^2}\right)\ .
\label{1prime}
\end{eqnarray}
Here we use $(r^nE)={\rm const.}$ and  $C_1(t)$ is a constant of integral.
Same as this, substituting (\ref{vector_subst}) into the second equation of (\ref{vector_EOM}) and integrating this with respect to $t$ reads 
\begin{eqnarray}
2(Er^n)C+C_2(r)=\frac{\kappa_v-(n-1)}{n-1}\phi-r^3T\left(\left(\frac{h_1}{r^2}\right)^{\prime}-\frac{{\dot h_2}}{r^2}\right)\ .
\label{2prime}
\end{eqnarray}
Comparison (\ref{1prime}) with (\ref{2prime}) shows that $C_1(t)=C_2(r)={\rm const.}$ and this constant can be absorbed into $\phi$. 
Therefore the three equations (\ref{vector_EOM}) are reduced into one equation
\begin{eqnarray}
2(Er^n)C=\frac{\kappa_v-(n-1)}{n-1}\phi-r^3T\left(\left(\frac{f}{r^2T^{\prime}}\phi^{\prime}\right)^{\prime}-\frac{1}{fr^2T^{\prime}}{\ddot \phi}\right)\ .
\label{vector_perturbation_gravity}
\end{eqnarray}
Same as this substitution, (\ref{vector_subst}) makes Maxwell equation (\ref{vector_maxwell_eq})  
\begin{eqnarray}
&\ &\frac{1}{f}{\ddot C}-\frac{1}{r^{n-2}}\partial_{r}\left(r^{n-2}f\partial_r C\right)\nonumber\\
&\ &\hspace{0.8cm}+\frac{E}{rT}\left(2(Er^n)C-\frac{\kappa_v-(n-1)}{n-1}\phi\right)
+\frac{\kappa_v+(n-1)}{r^2}C=0\ .
\label{vector_perturbation_maxwell}
\end{eqnarray}

As we have seen, $\phi$ determine the perturbation of gravitational field and $C$ does Maxwell field. 
Therefore, these are the master variables and (\ref{vector_perturbation_gravity}) and (\ref{vector_perturbation_maxwell}) are the master equations 
for vector type perturbations. 
 
Finally, we'd like to alter these two equations into a Scr${\rm \ddot o}$dinger equation with two components. 
To do so, we must change three points: Firstly, we change the normalization as 
\begin{eqnarray}
\Psi=\frac{\phi}{r\sqrt{T^{\prime}}},\quad \zeta=\sqrt{\frac{2(n-1)}{\kappa_v-(n-1)}}r^{(n-2)/2}C\ .
\end{eqnarray}
In this, we use the assumption that $T^{\prime}$ is always positive. 
Secondly, we switch radial coordinate $r$ to $r^*$. Finally,  we Fourier transforme like $\Psi\rightarrow\Psi e^{i\omega t}$ and $\zeta\rightarrow\zeta e^{i\omega t}$.  
Then, (\ref{vector_perturbation_gravity}) and (\ref{vector_perturbation_maxwell}) become a Scr${\rm \ddot o}$dinger equation with two components as 
\begin{eqnarray}
{\mathcal H}
\left(
\begin{array}{c}
\Psi\\
\zeta
\end{array}
\right)
=\omega^2
\left(
\begin{array}{c}
\Psi\\
\zeta
\end{array}
\right)\ ,
\label{vector_master_eq}
\end{eqnarray}
where
\begin{eqnarray}
{\mathcal H}=-\partial_{r^*}^2+\left(
\begin{array}{cc}
V_g(r)&V_c(r)\\
V_c(r)&V_{em}(r)
\end{array}
\right)
\end{eqnarray}
and
\begin{eqnarray}
V_g(r)&=&\frac{\kappa_v-(n-1)}{n-1}\frac{fT^{\prime}}{rT}+r\sqrt{T^{\prime}}\partial_{r^*}^2\frac{1}{r\sqrt{T^{\prime}}}\ ,\nonumber\\
V_c(r)&=&-2(Er^n) \sqrt{\frac{\kappa_v-(n-1)}{2(n-1)}}\frac{f\sqrt{T^{\prime}}}{r^{(n+2)/2}T}\ ,\nonumber\\
V_{em}(r)&=&(\kappa_v+(n-1))\frac{f}{r^2}+2(Er^n)^2\frac{f}{r^{n+1}T}+r^{-(n-2)/2}\partial_{r^*}^2r^{(n-2)/2}\ .
\end{eqnarray}
Note that these equations are not decomposed due to higher curvature collections. 
In Einstein limit, owing to $T(r)=r^{n-1}$, the above potential matrix can be diagonalized by constant eigenvectors. 
This indicates that our Schr${\ddot {\rm o}}$dinger equation can be decomposed into two equations by taking suitable linear combinations of $\psi$ and $\zeta$. 
In fact, we can do so by using the combination 
$\alpha_{\pm}\psi-\zeta$ 
with 
\begin{eqnarray}
\alpha_{\pm}=\frac{(n^2-1){\mathcal M}\pm\sqrt{(n^2-1)^2{\mathcal M}^2+8(\kappa_v-(n-1))(r^nE)^2}}{2(r^nE)\sqrt{2(\kappa_v-(n-1))}}, 
\end{eqnarray}
which is consistent with \cite{Kodama}. 
Against this, because $T(r)$ is more complicated in general Lovelock theory, 
we must consider the above coupling system.

\subsection{Stability Analysis}
In this subsection, we show that the Schr${\rm{\ddot o}}$dinger equation (\ref{vector_master_eq}) has no negative eigenvalue states.

In order to show this, for ${\vec \Psi}=(\Psi,\ \zeta)^T$, we define the inner product as 
\begin{eqnarray}
({\vec \Psi}_1,\ {\vec \Psi}_2)=\int_{-\infty}^{\infty}dr^{*}\left[\Psi_1^*\Psi_2+\zeta_1^*\zeta_2\right]\ .
\end{eqnarray}

Here, we prove that charged Lovelock black holes are stable for vector type perturbations.   
In order to show this, we prove ${\mathcal H}$ is an essentially positive-definite self-adjoint operator. 
For this, because  ${\mathcal H}$ with $C_0^{\infty}(r^*)\times C_0^{\infty}(r^*)$ is a symmetric operator,  
it is sufficient to check this operator is positive-definite~\cite{Kodama:2003jz}. 

We assume ${\vec \Psi_0}\in C_0^{\infty}(r^*)\times C_0^{\infty}(r^*)$, 
then $({\vec \Psi_0}, {\mathcal H}{\vec \Psi_0})$ can be estimated as 
\begin{eqnarray}
({\vec \Psi_0}, {\mathcal H}{\vec \Psi_0})&=&\int dr^*\Biggl[-\psi_0^*\partial_{r^*}^2\psi_0-\zeta_0^*\partial_{r^*}^2\zeta_0\nonumber\\
&\ &\hspace{1.5cm}+V_g|\psi_0|^2+V_c(\psi_0^*\zeta_0+\psi_0\zeta_0^*)+V_{em}|\zeta_0|^2\Biggr]\nonumber
\end{eqnarray}
\begin{eqnarray}
\ \ &=&\int  dr^*\Biggl[
\left|\partial_{r^*}\psi_0+\left(\partial_{r^{*}}\ln(r\sqrt{T^{\prime}})\right)\psi_0\right|^2
+\left|\partial_{r^*}\zeta_0-\frac{n-2}{2}\frac{f}{r}\zeta_0\right|^2\nonumber\\
&\ &\hspace{1.0cm}+\frac{\kappa_v-(n-1)}{n-1}\frac{fT^{\prime}}{rT}|\psi_0|^2
+2(Er^n)^2\frac{f}{r^{n+1}T}|\zeta_0|^2\nonumber\\
&\ &\hspace{1.5cm}+(\kappa_v+(n-1))\frac{f}{r^2}|\zeta_0|^2\nonumber\\
&\ &\hspace{2.0cm}-\sqrt{\frac{2(\kappa_v-(n-1))}{n-1}}(Er^n)\frac{f\sqrt{T^{\prime}}}{r^{(n+2)/2}T}(\psi_0^*\zeta_0+\psi_0\zeta_0^*)
\Biggr]\nonumber\\
\ \ \ &=&\int  dr^*\Biggl[
\left|\partial_{r^*}\psi_0+\left(\partial_{r^{*}}\ln(r\sqrt{T^{\prime}})\right)\psi_0\right|^2
+\left|\partial_{r^*}\zeta_0-\frac{n-2}{2}\frac{f}{r}\zeta_0\right|^2\nonumber\\
&\ &\hspace{1.5cm}+\frac{f}{rT}\left| \frac{\sqrt{2}(Er^n)}{r^{n/2}}\zeta_0-\sqrt{\frac{\kappa_v-(n-1)}{n-1}}\sqrt{T^{\prime}}\psi_0\right|^2\nonumber\\
&\ &\hspace{2.0cm}+(\kappa_v+(n-1))\frac{f}{r^2}|\zeta_0|^2
\Biggr]>0\ ,
\end{eqnarray}
where we use Gauss theorem in the second equality and neglect boundary terms because  $\Psi_0$ and $\zeta_0$ are in $C_0^{\infty}(r^*)$.  
This calculation shows that ${\mathcal H}$ with $C_0^{\infty}(r^*)\times C_0^{\infty}(r^*)$ is positive-definite. 
Then, since ${\mathcal H}$ with $C^\infty_0 \times C^\infty_0$ is essentially self-adjoint, ${\mathcal H}$ can be uniquely extended to a positive-definite self-adjoint operator, 
so there is no instability under vector type perturbations.  

\section{Scalar Type Perturbations}
\label{section5}

In this section, we derive the master equations and present conditions for the instability under scalar type perturbations. 
In this section, we assume ${\mathcal M}> {\mathcal M}_{ex}$ and  also assume that $T^{\prime}$ is always positive outside $r_{out}$. 

\subsection{Gravitational Perturbations}
Firstly, we consider the metric perturbations. In this paper, we take the Zerilli gauge in which metric perturbations are described as 
\begin{eqnarray}
\delta g_{\mu \nu}=
	\left(
	\begin{array}{cc|c}
	fH_0{\mathcal Y}&H_1{\mathcal Y}&0\\
	sym&H{\mathcal Y}/f&0\\ \hline
	sym& sym&r^2K{\mathcal Y} \gamma_{ij}
	\end{array}
	\right) \ ,
	\label{deltag_scalar}
\end{eqnarray}
where ${\mathcal Y}$ is the scalar harmonics which is characterized by the eigenequation ${\mathcal Y}_{|l}{}^{|l}=-\kappa_s {\mathcal Y}$ with $\kappa_s=\ell(\ell+n-1)$ and $\ell=0,1,2,\cdots$. 

For deriving the master equation, 
it is sufficient that we calculate $\delta {\mathcal G}_i^j$$(i\neq j)$,  $\delta {\mathcal G}_t^r$, $\delta {\mathcal G}_t^t$,  $\delta {\mathcal G}_r^i$ and $\delta {\mathcal G}_r^r$~\cite{Gleiser:2005ra}.  
By the above metric perturbations, we can derive the following results~\cite{Takahashi:2009dz}; 
\begin{eqnarray}
	 &\ &\delta {\mathcal G}_i{}^j=\frac{1}{2(n-1)r^n}(T^{\prime}H_0-T^{\prime}H-rT^{\prime\prime}K){\mathcal Y}^{|j}{}_{|i}\ ,\nonumber\\
	&\ &\delta {\mathcal G}_t{}^r=\frac{fT}{2r^{n+1}}\left[-\kappa_s H_1 +n\left\{\left(r-\frac{r^2f^{\prime}}{2f}\right)K+r^2K^{\prime}-rH\right\}^{\cdot}\right]{\mathcal Y}\ ,\nonumber\\
	&\ &\delta {\mathcal G}_t{}^t=\frac{1}{2r^{n+1}}\Biggl[\left\{-\kappa_s T-nr(fT)^{\prime}\right\}H-nrfTH^{\prime}+(n-\kappa_s)rT^{\prime}K\nonumber\\
	                     &\ &\hspace{3cm}+\left\{\frac{nr^2f^{\prime}T}{2}+nf(r^2T)^{\prime}\right\}K^{\prime}+nr^2fTK^{\prime\prime}\Biggr]{\mathcal Y}\ ,\nonumber\\
	&\ &\delta {\mathcal G}_r{}^i=\frac{1}{2r^{n+1}}\Biggl[\left(T^{\prime}+\frac{f^{\prime}T}{2f}\right)H-rT^{\prime}K^{\prime}+
	\left(\frac{f^{\prime}T}{2f}-\frac{T}{r}\right)H_0+TH_0^{\prime}-\frac{T}{f}{\dot H}_1\Biggr]{\mathcal Y}^{|i}\ , \nonumber\\
	&\ &\delta {\mathcal G}_r{}^r=\frac{1}{2r^{n+1}}\Biggl[2nrT{\dot H}_1-\frac{nr^2T}{f}{\ddot K}+(n-\kappa_s)rT^{\prime}K+\left(nr^2fT^{\prime}+\frac{nr^2f^{\prime}T}{2}\right)K^{\prime}\nonumber\\
	&\ &\hspace{3cm}-nr(fT)^{\prime}H+\kappa_sTH_0-nrfTH_0^{\prime}\Biggr]{\mathcal Y}\ . \label{scalar_delta_Lovelock}
\end{eqnarray} 
These are same as our previous calculation for neutral black holes except for the detail expression of $f(r)$. 

\subsection{Scalar Perturbations for Maxwell Field}
Next, we examine the scalar perturbations for Maxwell field. We start form  perturbations of the field strength which has the $U(1)$ gauge invariance. 
We describe this as 
\begin{eqnarray}
\delta F_{\mu\nu}=
\left(
\begin{array}{cc|c}
0&X {\mathcal Y}&Y{\mathcal Y}_{|i}\\
-X{\mathcal Y}&0&Z{\mathcal Y}_{|i}\\
\hline
\multicolumn{2}{c|}{{\bf anti\ sym.}}&0
\end{array}
\right)\ ,
\label{scalar_maxwell_ansatz}
\end{eqnarray}
where  ${\mathcal Y}$ is the scalar harmonics. Note that $\delta F_{ij}=0$ because we can not construct antisymmetric tensors  from scalar functions. 

The start point (\ref{scalar_maxwell_ansatz}) enables us to calculate the identity $\delta F_{[\mu\nu;\lambda]}=0$, 
Maxwell equations $\delta (F^{\mu\nu}{}_{;\nu})=0$ and the energy momentum tensor $\delta T_{\mu}{}^{\nu}$. 
Firstly, we check the identity. It is easy to show that the components other than $(\mu,\ \nu,\ \lambda)=(t,\ r,\ i)$ are trivial and the non-trivial component reads 
\begin{eqnarray}
X=Y^{\prime}-{\dot Z}\ .\label{scalar_part_identity}
\end{eqnarray}
Secondly, we calculate  Maxwell equations $\delta(F^{\mu\nu}{}_{;\nu})=0$. From $\mu=t$ component, we can gain a equation 
\begin{eqnarray}
\frac{1}{2}(r^nE)\partial_r(H_0-H+nK)-\partial_r(r^nX)+\frac{\kappa_sr^{n-2}}{f}Y=0\ .
\label{scalar_max_t}
\end{eqnarray}
Same as this, $\mu=r$ component reads 
\begin{eqnarray}
\frac{1}{2}(r^nE)\partial_t(H_0-H+nK)-\partial_t(r^nX)+\kappa_sr^{n-2}fZ=0
\label{scalar_max_r}
\end{eqnarray}
and $\mu=i$ components are 
\begin{eqnarray}
\partial_r(r^{n-2}fZ)-\partial_t\left(\frac{r^{n-2}}{f}Y\right)=0\ .
\label{scalar_max_i}
\end{eqnarray}

Before calculating $\delta T_{\mu}{}^{\nu}$, let us reduce the four equations (\ref{scalar_part_identity})$\sim$(\ref{scalar_max_i}) into one equation. 
From (\ref{scalar_max_i}), we can define a new variable $B$ as 
\begin{eqnarray}
Z=\frac{1}{r^{n-2}f}{\dot B},\quad Y=\frac{f}{r^{n-2}}B^{\prime}\ .\label{scalar_subst}
\end{eqnarray}
Substituting (\ref{scalar_subst}) into (\ref{scalar_max_t}) and integrating this with respect to $r$, (\ref{scalar_max_t}) becomes
\begin{eqnarray}
\frac{1}{2}(r^nE)(H_0-H+nK)-r^nX+\kappa_s B+C_1(t)=0\ ,\nonumber
\end{eqnarray}
where $C_1(t)$ is a constant of integral. We use $r^nE={\rm const.}$ in this integral. 
Same as this, (\ref{scalar_max_r}) becomes
\begin{eqnarray}
\frac{1}{2}(r^nE)(H_0-H+nK)-r^nX+\kappa_s B+C_2(r)=0\ .\nonumber
\end{eqnarray}
Here $C_2(r)$ is also a constant of integral. 
Then, a comparison with above two equations reads $C_1(t)=C_2(r)={\rm const.}$ and so this term can be absorbed into $B$. Therefore, 
\begin{eqnarray}
\frac{1}{2}(r^nE)(H_0-H+nK)-r^nX+\kappa_s B=0\ .\label{scalar_max_kore}
\end{eqnarray}
Eq.(\ref{scalar_subst}) and Eq.(\ref{scalar_max_kore}) suggest that $B$ (and gravitational perturbations) determines the perturbations of Maxwell field, 
so we can say $B$ is the master variable for Maxwell field. 
The evolution equation for the master variable $B$ can be derived  eliminating $X$ in (\ref{scalar_max_kore}) 
by (\ref{scalar_part_identity}) and (\ref{scalar_subst}). The result is 
\begin{eqnarray}
\frac{r^2}{f}\partial_t^2B-r^n\partial_r\frac{f}{r^{n-2}}\partial_rB+\kappa_sB+\frac{(r^nE)}{2}(H_0-H+nK)=0\ .
\label{scalar_maxwell_pert}
\end{eqnarray}

Finally, for the first order perturbations of Lovelock equations, we calculate the first order of the energy momentum tensor (\ref{Maxwell_EMtensor}). 
Eq.(\ref{Maxwell_EMtensor}), the background electric filed $E(r)$, metric perturbations (\ref{deltag_scalar}) and perturbations of Maxwell field (\ref{scalar_maxwell_ansatz}) yield
\begin{eqnarray}
&\ &\delta T_{t}{}^t=\delta T_r{}^r=\left[EX+\frac{1}{2}E^2(H-H_0)\right]{\mathcal Y}=\left[\frac{\kappa_s}{r^{n}}B+\frac{n}{2}EK\right]{\mathcal Y}\ ,\nonumber\\
&\ &\delta T_t{}^r=0,\quad\delta T_r{}^i=\frac{E}{r^2f}Y{\mathcal Y}^{|i}=\frac{E}{r^n}B^{\prime}{\mathcal Y}^{|i},\quad\delta T_i{}^j=0\ (i\neq j)\ . \label{scalar_EM_EM}
\end{eqnarray}
Note that we use the relation (\ref{scalar_max_kore}) in $(t,t),\ (r,r)$ components and also use the relation (\ref{scalar_subst}) in $(r,i)$ components. 

\subsection{Master Equations}

From the first order Lovelock tensor (\ref{scalar_delta_Lovelock}) and that of energy momentum tensor (\ref{scalar_EM_EM}), the components we concentrate on are
\begin{eqnarray}
&\ &T^{\prime}H_0-T^{\prime}H=rT^{\prime\prime}K\ ,
\label{scalar_ij_com}\\ 
&\ & -\kappa_sH_1+n\left\{\left(r-\frac{r^2f^{\prime}}{2f}\right)K+r^2K^{\prime}-rH\right\}^{\cdot}=0\ ,
\label{scalar_tr_com}\\
&\ &(n-\kappa_s)rT^{\prime}K+\left(\frac{nr^2f^{\prime}T}{2}+nf(r^2T)^{\prime}\right)K^{\prime}+nr^2fTK^{\prime\prime}\nonumber\\
&\ &\hspace{1cm} -(\kappa_s T+nr(fT)^{\prime})H-nrfTH^{\prime}=2r^{n+1}E\left(\frac{\kappa_s}{r^n}B+\frac{n}{2}EK\right)\ ,
\label{scalar_tt_com}\\
&\ &\left(T^{\prime}+\frac{f^{\prime}T}{2f}\right)H-rT^{\prime}K^{\prime}+
	\left(\frac{f^{\prime}T}{2f}-\frac{T}{r}\right)H_0+TH_0^{\prime}-\frac{T}{f}{\dot H}_1=2ErB^{\prime}\ ,
\label{scalar_ri_com}\\
&\ &(n-\kappa_s)rT^{\prime}K+\left(nr^2fT^{\prime}+\frac{nr^2f^{\prime}T}{2}\right)K^{\prime}-nr(fT)^{\prime}H\nonumber\\
&\ &\hspace{0.7cm} +\kappa_s TH_0-nrfTH_0^{\prime}+2nrT{\dot H}_1-\frac{nr^2T}{f}{\ddot K} =2r^{n+1}E\left(\frac{\kappa_s}{r^n}B+\frac{1}{2}nEK\right)\ .
\label{scalar_rr_com}
\end{eqnarray}

From now on, we'd like to construct the master equation.  
In order to derive, we must define the master variable $\phi$ and denote the gravitational perturbations by this $\phi$. 
Same as the analysis  for neutral black holes~\cite{Takahashi:2009dz}, 
we define the master variable $\phi$ as 
\begin{eqnarray}
H_1=\frac{r}{f}\left({\dot \phi}+{\dot K}\right)\ .
\label{scalar_master_variable_def}
\end{eqnarray}
Then, from (\ref{scalar_ij_com}), (\ref{scalar_tr_com}) and (\ref{scalar_tt_com}), we can express $H_0$,  $H$ and $K$ as follows~\cite{Takahashi:2009dz}; 
\begin{eqnarray}
H_0&=&H+\frac{rT^{\prime\prime}}{T^{\prime}}K\ ,\nonumber\\
H&=&-\frac{\kappa_s}{nf}\phi+rK^{\prime}-\frac{{\mathcal A}(r)}{2nf}K\ ,\nonumber\\
K&=&\frac{4nrfE}{{\mathcal A}T}B-\frac{2}{{\mathcal A}}\left[nrf\phi^{\prime}+\left(\kappa_s+nrf\frac{T^{\prime}}{T}\right)\phi\right]\ . 
\label{scalar_kannkeishiki}
\end{eqnarray}
where
\begin{eqnarray}
{\mathcal A}(r)=2\kappa_s+nrf^{\prime}-2nf\ .
\label{}
\end{eqnarray}
Above equations show that two variable $\phi$ and $B$ express the perturbative variables and so these are master variables. 
Therefore, we must construct evolution equations for these variables in order to examine the stability of background solution. 
That for $B$ has already been derived as (\ref{scalar_maxwell_pert}), but this include metric perturbations. 
Then, substituting (\ref{scalar_kannkeishiki}) into (\ref{scalar_maxwell_pert}) reads 
\begin{eqnarray}
&\ &{\ddot B}-f^2B^{\prime\prime}+f^2\left(\ln\left(\frac{r^{n-2}}{f}\right)\right)^{\prime}B^{\prime}+\frac{\kappa_sf}{r^2}B\nonumber\\
&\ &\ \ +\frac{(r^nE)f}{2r{\mathcal A}T}\left(\ln(r^nT^{\prime})\right)^{\prime}\times\left(4nrEfB-2nrfT\phi^{\prime}-2(\kappa_sT+nrfT^{\prime})\phi\right)=0\ .
\label{B_master}
\end{eqnarray}
The evolution equation for $\phi$ is derived from $nrf\times$(\ref{scalar_ri_com})$+$(\ref{scalar_rr_com}) with (\ref{scalar_kannkeishiki}) (see~\cite{Takahashi:2009dz}) and the result is 
\begin{eqnarray}
&\ &{\ddot \phi}-f^2\phi^{\prime\prime}+f^2\left(\ln\left(\frac{{\mathcal A}^2}{r^2fT^{\prime}}\right)\right)^{\prime}\phi^{\prime}\nonumber\\
&\ &\ \ +\frac{f}{nr^2T}\left[\left(2\frac{({\mathcal A}T)^{\prime}}{{\mathcal A}T}-\frac{T^{\prime\prime}}{T^{\prime}}\right)(\kappa_s rT+nr^2fT^{\prime})-n(r^2fT^{\prime})^{\prime}\right]\phi\nonumber\\
&\ &\ \ \ +\left[-f\frac{4\kappa_s(Er^n)}{nr^{n+1}T}+\frac{2f^2(Er^n)}{r^nT}\left(\ln\left(\frac{fT^{\prime}}{r^{n-2}({\mathcal A}T)^2}\right)\right)^{\prime}\right]B=0\ .
\label{phi_master}
\end{eqnarray}
These coupled two equations (\ref{B_master}) and (\ref{phi_master}) determine the behavior of $\phi$ and $B$.  
These two functions  determine all perturbative variables, so these two equations are the master equations.
 
Here, we derive  a Schr${\rm{\ddot o}}$dinger equation with two components like (\ref{vector_master_eq}) from these two equations. 
Against the case for vector type perturbations, it is more complicated  because there is $\phi^{\prime}$ in 
evolution equation for $B$. 
Therefore, in order to transform these equations into a Schr${\rm{\ddot o}}$dinger type equation, 
we must eliminate this $\phi^{\prime}$. 
For this, we must consider linear combinations of the master variables $\phi$ and $B$. 
For example, the following combinations are fit for our purpose; 
\begin{eqnarray}
\Psi\equiv\frac{r\sqrt{T^{\prime}}}{{\mathcal A}}\phi,\quad \zeta=\frac{1}{r^{(n-2)/2}\sqrt{n(\kappa_s-n)}}\left(B-\frac{n(r^nE)}{{\mathcal A}}\phi\right)\ .
\label{master_variables}
\end{eqnarray}
By using these variables and tortoise coordinate $r^*$,  we can obtain 
\begin{eqnarray}
{\mathcal H}
\left(
\begin{array}{c}
\Psi\\
\zeta
\end{array}
\right)
=\omega^2
\left(
\begin{array}{c}
\Psi\\
\zeta
\end{array}
\right)\ ,
\label{scalar_master_eq}
\end{eqnarray}
where
\begin{eqnarray}
{\mathcal H}=-\partial_{r^*}^2+\left(
\begin{array}{cc}
V_g(r)&V_c(r)\\
V_c(r)&V_{em}(r)
\end{array}
\right)\label{scalar_hamiltonian}
\end{eqnarray}
and
\begin{eqnarray}
V_g(r)&=&\kappa_s\frac{f}{nr}\left(4(\kappa_s-n)\frac{T^{\prime}}{{\mathcal A}T}-\frac{T^{\prime\prime}}{T^{\prime}}\right)\nonumber\\
&\ &\hspace{0.5cm}+\frac{2n(r^nE)^2f^2}{{\mathcal A}Tr^n}\left(\ln\left(\frac{fT^{\prime}}{r^{n-2}({\mathcal A}T)^2}\right)\right)^{\prime}+\frac{{\mathcal A}T}{r\sqrt{T^{\prime}}}f\partial_r\left(f\partial_r\frac{r\sqrt{T^{\prime}}}{{\mathcal A}T}\right)\ ,\nonumber\\
V_{c}(r)&=&\sqrt{\frac{\kappa_s-n}{n}}\frac{\sqrt{T^{\prime}}}{r^{n/2}{\mathcal A}}\left[-\kappa_s\frac{4(r^nE)f}{rT}+\frac{2n(r^nE)f^2}{T}\left(\ln\left(\frac{fT^{\prime}}{r^{n-2}({\mathcal A}T)^2}\right)\right)^{\prime}\right]\ ,\nonumber\\
V_{em}(r)&=&fr^{(n-2)/2}\partial_r(f\partial_r\frac{1}{r^{(n-2)/2}})\nonumber\\
&\ &\hspace{0.3cm}+\kappa_s\frac{f}{r^2}\left(1+\frac{4(Er^n)^2}{{\mathcal A}Tr^{n-1}}\right)+\frac{2n(r^nE)^2f^2}{r^n{\mathcal A}T}\left(\ln\left(\frac{r^{2n-2}({\mathcal A}T)^2}{f}\right)\right)^{\prime}\ .
\label{}
\end{eqnarray}
Here, we Fourier transforme like $\Psi\rightarrow \Psi e^{i\omega t}$ and $\zeta\rightarrow \zeta e^{i\omega t}$. 
Note that, in Einstein limit, we can decompose the above equation into two Schr${\ddot {\rm o}}$dinger equations by taking the linear combinations like 
$\psi-\alpha_{\pm}\zeta\ $
where
\begin{eqnarray}
\alpha_{\pm}=\frac{1}{4(r^nE)}\sqrt{\frac{n(n-1)}{\kappa_s-n}}(n+1)\left({\mathcal M}\pm\sqrt{{\mathcal M}^2+\frac{16(\kappa_s-n)(r^nE)^2}{n(n-1)(n+1)^2}}\right) \ .
\end{eqnarray}
This is consistent with Ref.\cite{Kodama}. We can do this because $T(r)=r^{n-1}$ in Einstein theory. 
Against this, because $T(r)$ is more complicated in general Lovelock theory, 
we must consider the above coupling system.  

\subsection{Condition for Instability} 
In this subsection, we show the criterion for the instability under scalar type perturbations. 

Here, we show that  ``if $2T^{\prime 2}-TT^{\prime\prime}$ takes negative values somewhere in $r>r_{out}$, 
charged Lovelock black holes have the instability" when $T^{\prime}$ is always 	positive. 
In order to show this, we here also define the inner product as 
\begin{eqnarray}
({\vec \Psi}_1,\ {\vec \Psi}_2)=\int_{-\infty}^{\infty}dr^{*}\left[\Psi_1^*\Psi_2+\zeta_1^*\zeta_2\right]\ ,
\end{eqnarray}
where ${\vec \Psi}=(\Psi,\ \zeta)^T$. 
For this proof, it is convenient to use the following inequality; for any test function ${\vec \Psi}_{test}\in C_0^{\infty}\times C_0^{\infty}$, 
 the lower bound of spectra for ${\mathcal H}$ with $C_0^{\infty}\times C_0^{\infty}$ satisfies  
\begin{eqnarray}
\omega_0^2\cdot ({\vec \Psi}_{test},\ {\vec \Psi}_{test})\leq({\vec \Psi}_{test},\ {\mathcal H}{\vec \Psi}_{test})\ .
\end{eqnarray}
This inequality suggests that there exist instabilities if we can find trial function ${\vec \Psi}_{test}$ which satisfies 
$({\vec \Psi}_{test},\ {\mathcal H}{\vec \Psi}_{test})<0$. 

We assume $2T^{\prime 2}-TT^{\prime\prime}$ has negative regions and 
define $I$ as a closed set in the region $2T^{\prime 2}-TT^{\prime\prime}<0$. 
Then, we choose a trial function as ${\vec \Psi}_{test}=(\Psi_0,\ 0)^T$ where 
$\Psi_0$ is a sufficiently smooth function with compact support on $I$. 
Using this test function, $({\vec \Psi}_{test},\ {\mathcal H}{\vec \Psi}_{test})$ is evaluated as 
\begin{eqnarray}
({\vec \Psi}_{test},\ {\mathcal H}{\vec \Psi}_{test})&=&\int_I dr^*\left[-\Psi_0^*\partial_{r^*}^2\Psi_0+V_g(r)|\Psi_0|^2\right]\nonumber\\
&=&\int_I dr^* \Biggl[\left|\partial_{r^*}\Psi_0+f\left(\partial_r\ln\frac{{\mathcal A}T}{r\sqrt{T^{\prime}}}\right)\Psi_0\right|^2\nonumber\\
&\ &\hspace{0.5cm}+2\kappa_s\frac{f}{nr}\left(2(\kappa_s-n)\frac{T^{\prime}}{{\mathcal A}T}-\frac{T^{\prime\prime}}{2T^{\prime}}\right)|\Psi_0|^2\nonumber\\
&\ &\hspace{0.8cm}+\frac{2n(r^nE)^2f^2}{{\mathcal A}Tr^n}\left(\ln\left(\frac{fT^{\prime}}{r^{n-2}({\mathcal A}T)^2}\right)\right)^{\prime}|\Psi_0|^2\Biggr]\ .
\label{scalar_evaluate}
\end{eqnarray}
Note that we neglect the boundary terms in second equality because $\Psi_0$ is zero at the boundary of $I$. 
Furthermore, using the relation 
\begin{eqnarray}
{\mathcal A}T=2(\kappa_s-n)T-nr^{n+2}M^{\prime}(r)\ ,\nonumber
\end{eqnarray}
the second line of the last equation can be evaluated as 
\begin{eqnarray}
&\ &\int _I dr^*\left[2\kappa_s\frac{f}{nr}\left(2(\kappa_s-n)\frac{T^{\prime}}{{\mathcal A}T}-\frac{T^{\prime\prime}}{2T^{\prime}}\right)|\Psi_0|^2\right]\nonumber\\
&=&\int _I dr^*\left[2\kappa_s\frac{f}{nr}\left(\frac{2(\kappa_s-n)}{2(\kappa_s-n)-nr^{n+2}\frac{M^{\prime}(r)}{T}}\frac{T^{\prime}}{T}-\frac{T^{\prime\prime}}{2T^{\prime}}\right)|\Psi_0|^2\right]\nonumber\\
&<&\int _I dr^*\left[2\kappa_s\frac{f}{nr}\left(\frac{T^{\prime}}{T}-\frac{T^{\prime\prime}}{2T^{\prime}}\right)|\Psi_0|^2\right]=\kappa_s\int _I dr^*\left[\frac{f}{nrTT^{\prime}}\left(2T^{\prime 2}-TT^{\prime\prime}\right)|\Psi_0|^2\right]\ ,
\end{eqnarray}
where we use the positivity of  $T$ and $T^{\prime}$ and also use eq.(\ref{M_prime}) in the inequality. 
Then $({\vec \Psi}_{test},\ {\mathcal H}{\vec \Psi}_{test})$ satisfies the following inequality;
\begin{eqnarray}
({\vec \Psi}_{test},\ {\mathcal H}{\vec \Psi}_{test})&<&
\int_I dr^* \Biggl[\left|\partial_{r^*}\Psi_0+f\left(\partial_r\ln\frac{{\mathcal A}T}{r\sqrt{T^{\prime}}}\right)\Psi_0\right|^2\nonumber\\
&\ &\hspace{1.5cm}+\frac{2n(r^nE)^2f^2}{{\mathcal A}Tr^n}\left(\ln\left(\frac{fT^{\prime}}{r^{n-2}({\mathcal A}T)^2}\right)\right)^{\prime}|\Psi_0|^2\Biggr]\nonumber\\
&\ &\hspace{0.3cm}+\kappa_s\int _I dr^*\left[\frac{f}{nrTT^{\prime}}\left(2T^{\prime 2}-TT^{\prime\prime}\right)|\Psi_0|^2\right]\ .
\label{scalar_evaluate2}
\end{eqnarray}
In this equation, the second integral must be negative under our assumptions, 
therefore the second integral of (\ref{scalar_evaluate2}) tends to $-\infty$ when $\kappa_s=\ell(\ell+n-1)\rightarrow \infty$. 
Against this results, the first integral of (\ref{scalar_evaluate2}) converge when $\ell\rightarrow \infty$ like 
\begin{eqnarray}
&\ &\int_I dr^* \Biggl[\left|\partial_{r^*}\Psi_0+f\left(\partial_r\ln\frac{{\mathcal A}T}{r\sqrt{T^{\prime}}}\right)\Psi_0\right|^2
+\frac{2n(r^nE)^2f^2}{{\mathcal A}Tr^n}\left(\ln\left(\frac{fT^{\prime}}{r^{n-2}({\mathcal A}T)^2}\right)\right)^{\prime}|\Psi_0|^2\Biggr]\nonumber\\
&\ &\hspace{1cm}\rightarrow \int_I dr^* \Biggl[\left|\partial_{r^*}\Psi_0+f\left(\partial_r\ln\frac{T}{r\sqrt{T^{\prime}}}\right)\Psi_0\right|^2\Biggr]\ .
\end{eqnarray}
This is because integrands converge uniformly on $I$ like 
\begin{eqnarray}
&\ &\partial_{r^*}\Psi_0+f\left(\partial_r\ln\frac{{\mathcal A}T}{r\sqrt{T^{\prime}}}\right)\Psi_0\rightarrow \partial_{r^*}\Psi_0+f\left(\partial_r\ln\frac{T}{r\sqrt{T^{\prime}}}\right)\Psi_0\ ,\nonumber\\
&\ &\frac{2n(r^nE)^2f^2}{{\mathcal A}Tr^n}\left(\ln\left(\frac{fT^{\prime}}{r^{n-2}({\mathcal A}T)^2}\right)\right)^{\prime}|\Psi_0|^2\rightarrow 0\ ,
\end{eqnarray}
where these uniform convergences are supported by the theorem that continuity functions on closed set have maximum value and minimum value. 
Therefore, summarizing these results, we can denote that the r.h.s of (\ref{scalar_evaluate2}) tends to $-\infty$ as $\ell\rightarrow \infty$. 
This means that the lower bound of the spectra is negative  in sufficiently large $\ell$ modes and so background solution has instability for these modes. 

We'd like to summarize this section. We assume $T^{\prime}>0$ in $r>r_{out}$. Under this assumption, we derived master equations. 
We can unify these equations as a Schr${\ddot {\rm o}}$dinger equation with two components. 
We show that this  Schr${\ddot {\rm o}}$dinger operator has negative spectra when $2T^{\prime 2}-TT^{\prime\prime}$ has negative region. 
Therefore, we can denote that $2T^{\prime 2}-TT^{\prime\prime}$ is crucial for the stability of charged Lovelock black holes. 
This criteria is same as neutral case~\cite{Takahashi:2009dz}.

We have not shown the inverse statement so far. Then, even if $2T^{\prime 2}-TT^{\prime\prime}$ is always positive, we can not declare 
this black hole is stable. For example, in Einstein case, $T(r)$ is $r^{n-1}$, so  $2T^{\prime 2}-TT^{\prime\prime}=n(n-1)r^{2(n-2)}>0$. 
Then we can not say anything for Einstein case. 
Same as this, in 6-dimensions, this function can be evaluated as 
\begin{eqnarray}
2T^{\prime 2}-TT^{\prime\prime}=3\frac{r^3(\beta_2{\mathcal M}-2r^5)^2+8\beta_2{\mathcal Q}^2(\beta_2{\mathcal M}+3r^5)}{r^9(1+2\beta_2M(r))}>0\ ,
\label{6D_scalar_check}
\end{eqnarray}
so we cannot say anything for scalar perturbations in 6-dimensions .

\section{Numerical Results} 
\label{section6}

In section \ref{section3} and section \ref{section5}, 
we have shown that the behavior of $T(r)$ is crucial for instability. 
In detail, $T^{\prime}$ is critical for ghost, $T^{\prime\prime}$ is for tensor perturbations and $2T^{\prime 2}-TT^{\prime\prime}$ is for scalar perturbations. 
In this section, we check the behavior of these functions. 
For neutral cases, we can examine analytically because we can reduce these functions into the polynomial functions of $\psi$~\cite{Takahashi:2009dz}.  
However, such reduction can not be performed in charged case. Therefore, we numerically check 
the behaviors of  $T^{\prime}$, $T^{\prime\prime}$ and $2T^{\prime 2}-TT^{\prime}$ for various $(|{\mathcal Q}|,{\mathcal M})$ with some Lovelock couplings. 

In numerical calculation, we must use dimensionless parameters. 
So far we have discussed in the unit $\beta_1=1$; in this unit, we can not fix the scale of length. 
Same as our previous paper~\cite{Takahashi:2011qda}, we here also use $\beta_2$ for fixing this scale. 
This constant has a dimension of length squared. 
Therefore, we can relate $r$, $\psi$, ${\mathcal M}$, ${\mathcal Q}$ and Lovelock couplings $\beta_m$s to the dimensionless parameters as follows;  
\begin{eqnarray}
{\tilde r}\equiv r/\sqrt{\beta_2},\quad {\tilde \psi}\equiv \beta_2\psi,\quad \mu \equiv\beta_2^{-(n-1)/2}{\mathcal M},\quad Q\equiv\beta_2^{-(n-1)/2}{\mathcal Q}, \quad 
c_m\equiv\frac{\beta_m}{\beta_2^{m-1}}\ .\nonumber
\end{eqnarray}

Hereafter, we show some results for $5\sim 8$ dimensions.
The strategy of our numerical calculation is basically same as   our previous paper except for checking $2T^{\prime2}-TT^{\prime\prime}$~\cite{Takahashi:2011qda}. 
Note that $\mu_{ex}$ is the dimensionless extremal mass parameter which can be calculated from (\ref{determinant_extreme_mass}). 
$\mu_{tensor}$ is a border between stable and unstable for tensor type perturbations 
and $\mu_{scalar}$ corresponds to that for scalar type perturbations.  

\subsection{5-dimensions}

As we have mentioned above,  we use $\beta_2$ for fixing the scale of length. 
Then we need not regard the Lovelock couplings in 5-dimensions.

\begin{figure}[htbp]
  \begin{center}
   \includegraphics[width=90mm]{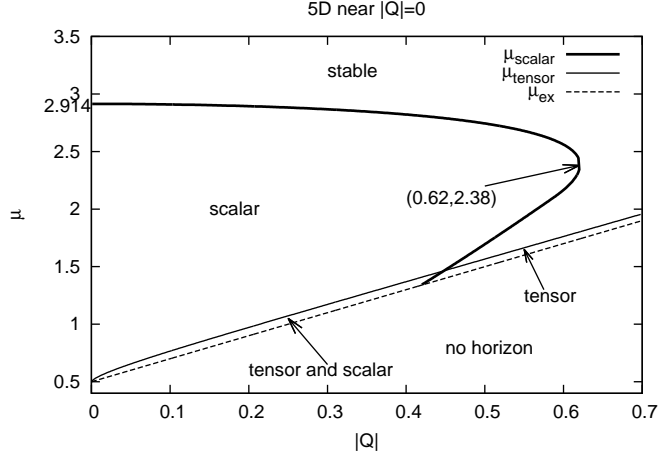}
  \end{center}
  \caption{Numerical results in 5-dimensions. 
  The scalar-unstable region localizes near the $\mu$-axis and there exists a slight gap between $\mu_{tensor}$ and $\mu_{ex}$. 
  This gap is about $O(10^{-1})$.
  Notice that ``stable" means both $T^{\prime\prime}$ and $2T^{\prime 2}-TT^{\prime\prime}$ have no negative region. 
  This is same for the following figures.}
  \label{fig5D}
  \end{figure}

We present the numerical results for 5-dimensions in Fig.\ref{fig5D}.  
For this figure, we check the region where $\mu_{ex}(|Q|)\sim \mu_{ex}(|Q|)+3$ for each $|Q|$ and 
the mesh size is $d\mu=dQ=10^{-3}$.  

As shown in our previous paper~\cite{Takahashi:2009dz}, the tensor-unstable region lies thinly on the extremal mass $\mu_{ex}(|Q|)$  
and $\mu_{tensor}(|Q|)$ converges to $0.5$ when $|Q|\rightarrow 0$. In this limit, $\mu_{ex}$ also converges to this value.  
Note that this thin region is over at $|Q|\sim 3$ which has been checked in our previous analysis. 
Against the tensor-unstable region, for scalar type perturbations, 
the unstable mass range is relatively wide and this unstable region localizes near the $\mu$-axis;  
the upper line of this region  is approximately expressed as $\mu\sim 2.914$ and this region suddenly disappear near $|Q|\sim 0.62$. 
For the ghost instability,  we have already checked the positivity of $T^{\prime}$ in (\ref{EGB_ghost}). 

Then, roughly speaking, charged Lovelock black hole with $\mu<2.914$  has the instability for scalar modes when $|Q|<0.62$; if $0.62<|Q|<3$, 
nearly extremal black hole is unstable for tensor type perturbations. We cannot detect the instability when black hole has more charge.  

\subsection{6-dimensions}

In 6-dimensions, same as 5-dimensional case, we need not alter Lovelock coefficients.  

As we have checked in section \ref{section3}, $T^{\prime}$ is always positive so there is no ghost. 
It has also mentioned in section \ref{section5} that  $2T^{\prime 2}-TT^{\prime\prime}$ is positive definite in 6-dimensions, 
which means we can not find the instability for scalar type perturbations. 
Then we can only detect the instability under tensor perturbations  and 
the results  are same as our previous analysis~\cite{Takahashi:2011qda};   
same as the tensor-unstable region in 5-dimensions, this region exists just on $\mu_{ex}(|Q|)$ and disappears at $|Q|\sim 3.28$. 
However, there also exists a difference; $\mu_{tensor}$ converges to $0.27$ in $|Q|\rightarrow 0$ while $\mu_{ex}\rightarrow 0$. 
Therefore, we can say that black hole with $\mu\sim\mu_{ex}$ has the instability under tensor type perturbations 
when $0\leq |Q|<3.28$. Especially, black hole has the instability also in neutral case against the 5-dimensional case.  

 \begin{figure}[htbp]
  \begin{center}
   \includegraphics[width=90mm]{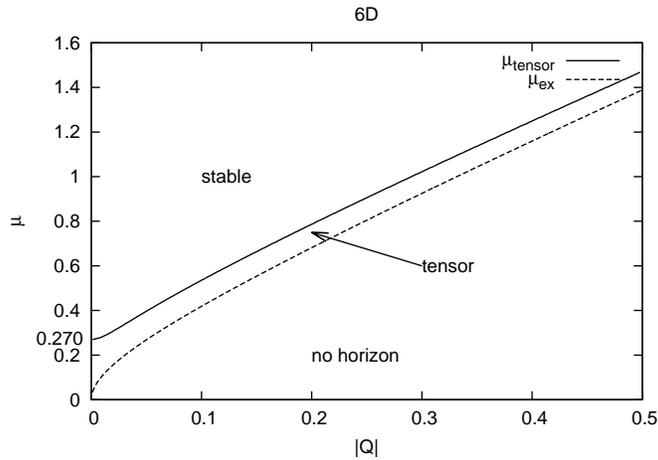}
  \end{center}
  \caption{Numerical results in 6-dimensions. 
  Unlike 5-dimensional case, there are no scalar-unstable regions due to eq.(\ref{6D_scalar_check}). 
  The tensor-unstable region slightly lies on the extreme line $\mu_{ex}(|Q|)$. 
  This region ends off at $|Q|\sim 3.28$. 
  Becasue $\mu_{ex}\rightarrow 0$ as $|Q|\rightarrow 0$, there also exists a tensor-unstable mass range in neutral case. }
  \label{fig6D}
  \end{figure}

\subsection{7-dimensions}

\begin{figure}[htbp]
 \begin{center}
  \includegraphics[width=90mm]{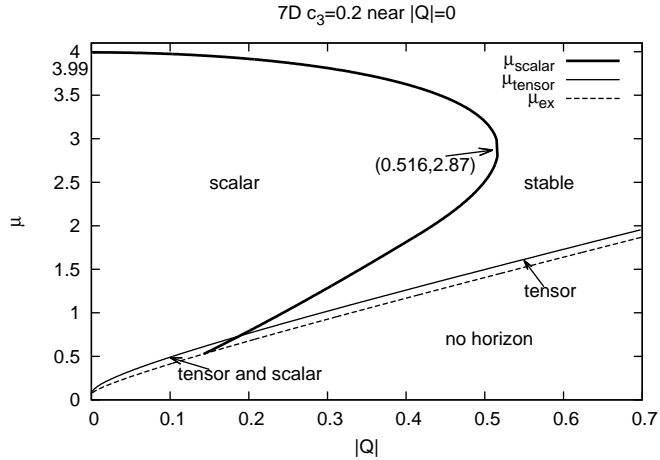}
 \end{center}
 \caption{Numerical results for 7-dimensions.
 We calculate this with $c_3=0.2$.  
  Both $\mu_{tensor}$ and $\mu_{ex}$ converge to $0.667$ as $Q\rightarrow 0$. Then there is no instability under tensor perturbations in neutral case. 
 Note that there exists no ghost region in $c_2=0.2$. 
 The appearances of the diagrams do not strongly depend on $c_3$ except for ghost regions. }
 \label{fig7D}
\end{figure}

We present the numerical results for 7-dimensions in Fig.\ref{fig7D}. 
This figure is calculated with $c_3=0.2$ in 7-dimensions. 
For this figure, we check  the region where $\mu_{ex}(|Q|)\sim \mu_{ex}(|Q|)+4.2$ for each $|Q|$ and 
the mesh size is $d\mu=dQ=10^{-3}$. 

This diagram is  almost same as 5-dimensional cases; 
the tensor-unstable region clings to $\mu_{ex}(|Q|)$ and the scalar one does to $\mu$-axis. 
Our previous analysis shows that the tensor-unstable region exists in $0<|Q|<4.695$~\cite{Takahashi:2011qda}. 
The upper bound of the scalar-unstable region is $\mu\sim 3.99$ and this region vanishes at $|Q|\sim 0.516$. 
When $c_3$ changes, the upper bound etc. change but the appearances of these unstable regions do not change.

When $c_3=0.2$, there are no ghost regions. However, when $c_3$ is larger than $0.25$, 
we can find ghost region near the origin of the diagram (see Fig.14 of our previous paper~\cite{Takahashi:2011qda}). 
 
Therefore, same as 5-dimensions, we can roughly say that 
black hole suffers from the instability under scalar perturbations if $Q\sim 0$ and $\mu$ is smaller than a certain value and 
 has instability under tensor modes when $\mu$ is as small as $\mu_{ex}$. 
Against 5-dimensional case, there exists $c_3$ dependance for ghost regions.    

\subsection{8-dimensions}

\begin{figure}[htbp]
 \begin{center}
  \includegraphics[width=90mm]{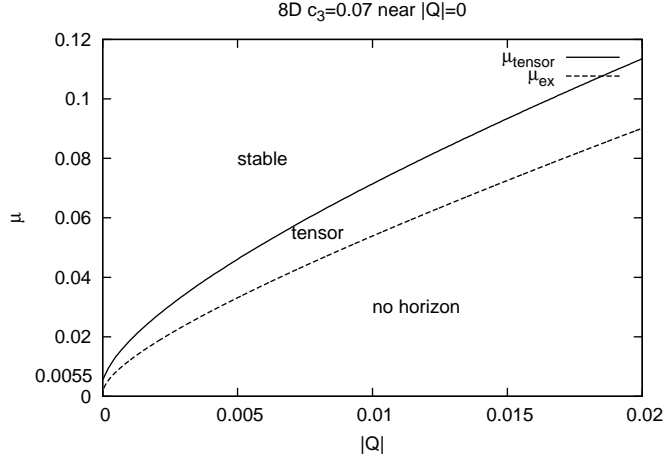}
 \end{center}
 \caption{Numerical results for 8-dimensions with $c_3=0.07$.  The appearance of this figure is very similar to the result for 6-dimensions; 
 There only exists the tensor-unstable region near extreme line $\mu_{ex}(|Q|)$. 
 $\mu_{tensor}$ converges to $0.0055$ and $\mu_{ex}$ tends to $0$ when $|Q|\rightarrow 0$. 
 Then there also exists the instability under tensor type perturbations when black holes are neutral.
 We cannot find ghost regions and scalar-unstable regions in $c_3=0.07$. }
 \label{fig8D_1}
\end{figure}

We present the two figures for 8-dimensions. 
Fig.\ref{fig8D_1} is the numerical result for $c_3=0.07$; 
we calculate the region in $\mu_{ex}(|Q|)<\mu<\mu_{ex}(|Q|)+0.05$ and the mesh size is $d\mu=dQ=10^{-4}$.  
This figure is almost same as that for 6-dimensional case. 
In this figure, there are no parameters which make $2T^{\prime 2}-TT^{\prime\prime}$ negative, so 
we can only find the instability under tensor type perturbations. 
For tensor type perturbations, the unstable region slightly lies on $\mu_{ex}(|Q|)$ and 
there also exists a gap between $\mu_{tensor}$ and $\mu_{ex}$ in $|Q|\rightarrow 0$.  
These properties are similar to 6-dimensional case.  
 
\begin{figure}
 \begin{center}
  \includegraphics[width=90mm]{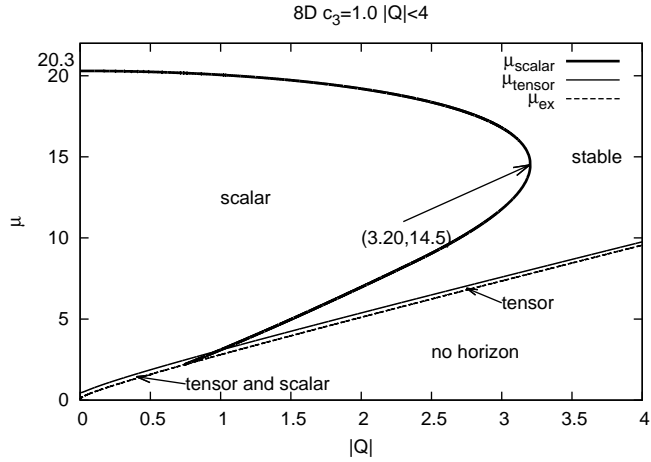}
 \end{center}
 \caption{Numerical results for 8-dimensions with  $c_3=1$. Against $c_3=0.07$ case, there exist a scalar-unstable region near $|Q|=0$. 
 Then, in 8-dimensions, which type instability black holes suffer from is very sensitive to third order Lovelock coupling $c_3$.
 In this figure, $\mu_{ex}\rightarrow 0$ and $\mu_{tensor}\rightarrow 0.459$ as $|Q|\rightarrow 0$.  
  Note that we cannot find ghost regions in $c_3=1$.}
 \label{fig8D_2}
\end{figure}

Against the above results, 
there exist  a scalar-unstable region in Fig.\ref{fig8D_2}.  
This figure is calculated with $c_3=1$ and check the region in $\mu_{ex}(|Q|)<\mu<\mu_{ex}(|Q|)+30$ with the mesh size  $d\mu=dQ=10^{-3}$. 
This figure is very similar to 5-dimensional diagram; 
a tensor-unstable region exists just on extreme line $\mu_{ex}(|Q|)$ and 
a scalar-unstable region localizes near $\mu$-axis. Furthermore, there is no ghost region in $c_3=1$. 

In Fig.\ref{fig8D_1} and Fig.\ref{fig8D_2}, there are no ghost regions. 
However, when $c_3$ becomes larger than 5.92, a ghost region appears near the origin of the diagram (see Fig.18 of \cite{Takahashi:2011qda}). 
Then, the appearance is similar to 7-dimensional case with sufficiently large $c_3$.

In 8-dimensions, as we have presented,  the appearance of diagrams are very responsive to the Lovelock coupling $c_3$. 
When $c_3$ is very small, this is same as 6-dimensional diagram. 
As $c_3$ becomes larger, the looks of the diagrams change from 
5-dimensional results  to 7-dimensional diagrams with large $c_3$.  
It is still an open issue why such dramatical changes occur in 8-dimensions. 

\subsection{Summary of Numerical Results}

In this section, we have numerically checked the condition for the instability and ghost. 
Here, we summarize the results.  

We numerically examine the behavior of $T(r)$ for various Lovelock coupling in 5, 6, 7 and 8 dimensions and 
plot the results in $Q-\mu$ diagrams.
From these results, we can read some common properties. 
The first is locations of unstable regions. 
The unstable regions for tensor type perturbations slightly lie on extreme line $\mu_{ex}$. 
The regions for scalar type perturbation, if exists,  localize at the $\mu$-axis and extend to the relatively large $\mu$. 
The second is manner of tensor-unstable region near $|Q|=0$ when there is no scalar-unstable region. 
For this case, in our numerical calculation, there must exist slight mass range in which black holes have the instability under tensor perturbations. 
 Therefore, whether scalar-unstable region exists or not, when black holes with nearly extreme mass have slight charge, 
they must be unstable; which type instability they have depends on parameters and dimensions, 
but they have at least one type instability. 

These results also lead some open questions. The first is dimensionality.  
The diagram for 5-dimensional case and that for 7-dimensional case are very similar:  
the result  for 6-dimensions and that for 8-dimension with small $c_3$ are alike. 
These remind us that the behaviors of black holes against perturbations are different  in even dimensions and odd dimensions. 
However, we do not have the answer for such a dimensionality. 
The second is the response to variation of Lovelock couplings. 
In 8-dimensions, for example, the appearance of diagram is sensitive to third order coupling $c_3$. 
These are  still open questions whether this is proper to 8 dimension or not and 
what causes such peculiarity.

\section{Conclusion}
\label{section7}

We have studied the stability of  charged Lovelock black hole solutions. 
We have derived master equation for vector type perturbations and scalar type perturbations. 
These are the Schr${\ddot {\rm o}}$dinger equations with two components. 
For vector type perturbations,  we have shown that the Schr${\ddot {\rm o}}$dinger operator for this type perturbations 
is an essentially positive definite self adjoint operator. Then charged Lovelock black holes are stable under vector type perturbations. 
On scalar type perturbations, we have presented the condition for instability. 
In detail, if $2T^{\prime 2}-TT^{\prime\prime}$ has negative regions, charged Lovelock black holes are unstable under scalar type perturbations. 
For tensor type perturbations, we have already shown that $T^{\prime}$ and $T^{\prime\prime}$ are crucial for stability. 
By numerically checking these criteria,  
for example in the 2nd order and the 3rd order Lovelock theory,  
we have shown that there exists  an unstable parameters;  
nearly extremal black holes have the instability under tensor type perturbations, and 
black holes with small charge have the instability under scalar type perturbations even if black holes have relatively large mass. 

One of future works is the exploration of more general conditions for stability under scalar type perturbations. 
In this paper, we have shown that black holes are unstable if $2T^{\prime 2}-TT^{\prime\prime}$ has negative region. 
However, the inverse statement has not been proved. 
Hence, so far, we can not say anything when $2T^{\prime 2}-TT^{\prime\prime}$ is always positive. 
Furthermore, by this criterion, we can not detect the instability under scalar type perturbation in Einstein theory~\cite{Konoplya:2007jv}. 
In this sense, it is interesting to find out more general conditions. 

It is interesting to investigate the relation between dynamical instability we have shown in this paper and thermodynamics. 
On the thermodynamics for Lovelock black holes, 
variation of Lovelock coefficients is also examined in Ref.~\cite{Kastor:2010gq}. 
In our paper, we have found that the appearances of $Q-\mu$ diagrams change dramatically as Lovelock coupling $c_3$ varies in 8 dimensions. 
Then it is interesting if such dramatical changes are found in the thermodynamics. 

The relation between instability and gravitational collapses might be important. 
In Lovelock theory, collapses of dust clouds have been examined~\cite{Maeda:2005ci}.   
Furthermore, these are extended to charged dust clouds~\cite{seiju}.  
In these, dependence of dimensions are found for, for example, naked singularity formations. 
Our results also depends dimensions, so there may exit relations between the instability and the gravitational collapse in Lovelock gravity. 
In dust collapses, the authors of above papers  have also pointed out  the tendency that 
higher curvature collections suppress formations of apparent horizons. 
These results  should express  that higher curvature collections make attractive force  weaker, and 
this property might be related to the instability of Lovelock black holes we discussed in this paper.

\begin{acknowledgements}
The author would like to thank Akihiro Ishibashi, Jiro Soda and Seiju Ohashi for useful comments and fruitful discussions.
This work is supported by the Japan Society for the Promotion of Science
(JSPS) grant No. 23 - 661 and  the Grant-in-Aid for the Global COE Program 
``The Next Generation of Physics, Spun from Universality and Emergence" 
from the Ministry of Education, Culture, Sports, Science and Technology (MEXT) of Japan.
\end{acknowledgements}


\begin{thebibliography}{99}
\bibitem{Giddings:2001bu}
  S.~B.~Giddings and S.~D.~Thomas,
  Phys.\ Rev.\  D {\bf 65}, 056010 (2002)
  [arXiv:hep-ph/0106219];
  S.~B.~Giddings and M.~L.~Mangano,
  Phys.\ Rev.\  D {\bf 78}, 035009 (2008)
  [arXiv:0806.3381 [hep-ph]].

\bibitem{Tangherlini:1963bw} 
  F.~R.~Tangherlini,
  Nuovo Cim.\  {\bf 27}, 636 (1963).

\bibitem{Myers:1986un} 
  R.~C.~Myers and M.~J.~Perry,
  Annals Phys.\  {\bf 172}, 304 (1986).
  
\bibitem{Emparan:2001wk} 
  R.~Emparan and H.~S.~Reall,
  Phys.\ Rev.\ D {\bf 65}, 084025 (2002)
  [hep-th/0110258].
  
\bibitem{Iguchi:2007is} 
  H.~Iguchi and T.~Mishima,
  Phys.\ Rev.\ D {\bf 75}, 064018 (2007)
  [Erratum-ibid.\ D {\bf 78}, 069903 (2008)]
  [hep-th/0701043].

\bibitem{Elvang:2007rd} 
  H.~Elvang and P.~Figueras,
  JHEP {\bf 0705}, 050 (2007)
  [hep-th/0701035].

\bibitem{Kodama:2003jz}
  H.~Kodama and A.~Ishibashi,
  Prog.\ Theor.\ Phys.\  {\bf 110}, 701 (2003)
  [arXiv:hep-th/0305147];\\
  A.~Ishibashi and H.~Kodama,
  Prog.\ Theor.\ Phys.\  {\bf 110}, 901 (2003)
  [arXiv:hep-th/0305185];\\
A.~Ishibashi and H.~Kodama,
  Prog.\ Theor.\ Phys.\ Suppl.\  {\bf 189}, 165 (2011)
  [arXiv:1103.6148 [hep-th]].
  
\bibitem{Kodama}
  H.~Kodama and A.~Ishibashi,
  Prog.\ Theor.\ Phys.\  {\bf 111}, 29 (2004)
  [arXiv:hep-th/0308128].

  
\bibitem{Konoplya:2007jv}
  R.~A.~Konoplya and A.~Zhidenko,
  Nucl.\ Phys.\  B {\bf 777}, 182 (2007)
  [arXiv:hep-th/0703231].

\bibitem{Dias:2009iu} 
  O.~J.~C.~Dias, P.~Figueras, R.~Monteiro, J.~E.~Santos and R.~Emparan,
  Phys.\ Rev.\ D {\bf 80}, 111701 (2009)
  [arXiv:0907.2248 [hep-th]];
  
  O.~J.~C.~Dias, P.~Figueras, R.~Monteiro and J.~E.~Santos,
  Phys.\ Rev.\ D {\bf 82}, 104025 (2010)
  [arXiv:1006.1904 [hep-th]];
  
  K.~Murata,
  Prog.\ Theor.\ Phys.\ Suppl.\  {\bf 189}, 210 (2011).


\bibitem{Tanahashi:2012si} 
  N.~Tanahashi and K.~Murata,
  arXiv:1208.0981 [hep-th].


\bibitem{Figueras:2011he} 
  P.~Figueras, K.~Murata and H.~S.~Reall,
  Class.\ Quant.\ Grav.\  {\bf 28}, 225030 (2011)
  [arXiv:1107.5785 [gr-qc]].

\bibitem{Lovelock:1972vz} 
  D.~Lovelock,
  J.\ Math.\ Phys.\  {\bf 13}, 874 (1972).
  
\bibitem{Lovelock:1971yv}
  D.~Lovelock,
  J.\ Math.\ Phys.\  {\bf 12} (1971) 498.
  
  
\bibitem{Christos} 
  C.~Charmousis,
  Lect.\ Notes Phys.\  {\bf 769}, 299 (2009)
  [arXiv:0805.0568 [gr-qc]];\\
  C.~Garraffo and G.~Giribet,
  Mod.\ Phys.\ Lett.\  A {\bf 23}, 1801 (2008)
  [arXiv:0805.3575 [gr-qc]].

\bibitem{Wheeler:1985nh}
  J.~T.~Wheeler,
  Nucl.\ Phys.\  B {\bf 273}, 732 (1986).

\bibitem{Dotti:2004sh}
  G.~Dotti and R.~J.~Gleiser,
  Class.\ Quant.\ Grav.\  {\bf 22}, L1 (2005)
  [arXiv:gr-qc/0409005];\\
  G.~Dotti and R.~J.~Gleiser,
  Phys.\ Rev.\  D {\bf 72}, 044018 (2005)
  [arXiv:gr-qc/0503117].

\bibitem{Gleiser:2005ra}
  R.~J.~Gleiser and G.~Dotti,
  Phys.\ Rev.\  D {\bf 72}, 124002 (2005)
  [arXiv:gr-qc/0510069];\\
  M.~Beroiz, G.~Dotti and R.~J.~Gleiser,
  Phys.\ Rev.\  D {\bf 76}, 024012 (2007)
  [arXiv:hep-th/0703074];\\
  R.~A.~Konoplya and A.~Zhidenko,
  Phys.\ Rev.\  D {\bf 77}, 104004 (2008)
  [arXiv:0802.0267 [hep-th]].


\bibitem{Takahashi:2009dz}
    T.~Takahashi and J.~Soda,
  Prog.\ Theor.\ Phys.\  {\bf 124}, 911 (2010)
  [arXiv:1008.1385 [gr-qc]];\\
  T.~Takahashi and J.~Soda,
  Prog.\ Theor.\ Phys.\  {\bf 124}, 711 (2010)
  [arXiv:1008.1618 [gr-qc]].
  
\bibitem{Takahashi:2011qda} 
  T.~Takahashi,
  Prog.\ Theor.\ Phys.\  {\bf 125}, 1289 (2011)
  [arXiv:1102.1785 [gr-qc]].

\bibitem{Myers:1988ze}
  R.~C.~Myers and J.~Z.~Simon,
  Phys.\ Rev.\  D {\bf 38}, 2434 (1988).
  
\bibitem{Kofinas:2007ns}
  G.~Kofinas and R.~Olea,
  JHEP {\bf 0711}, 069 (2007)
  [arXiv:0708.0782 [hep-th]].
  

\bibitem{Kastor:2010gq} 
  D.~Kastor, S.~Ray and J.~Traschen,
  Class.\ Quant.\ Grav.\  {\bf 27}, 235014 (2010)
  [arXiv:1005.5053 [hep-th]].
 
\bibitem{Maeda:2005ci}
  H.~Maeda,
  Class.\ Quant.\ Grav.\  {\bf 23}, 2155 (2006)
  [arXiv:gr-qc/0504028];\\
  S.~Ohashi, T.~Shiromizu and S.~Jhingan,
  Phys.\ Rev.\ D {\bf 84}, 024021 (2011)
  [arXiv:1103.3826 [gr-qc]].
  
  \bibitem{seiju}
S.~Ohashi, T.~Shiromizu and S.~Jhingan,
  Phys.\ Rev.\ D {\bf 86}, 044008 (2012)
  [arXiv:1205.5363 [gr-qc]].

\end{thebibliography}
\end{document}